\documentclass[reprint,superscriptaddress,showpacs,amsmath,amssymb,aip,jcp]{revtex4-1}

\usepackage{dcolumn,bm,book tabs,comment,natbib}
\usepackage[pdftex]{graphicx}
\usepackage[usenames,dvipsnames]{color}
\usepackage{colortbl,longtable}

\definecolor{URLCOL}{rgb}{0,0.52,0.83} 
\definecolor{LINKCOL}{rgb}{0.05,0.5,0} 
\definecolor{CITECOL}{rgb}{0.25,0,0.48} 
\definecolor{PREPRINTCOL}{rgb}{0.0,0.0,0.0} 

\usepackage[pdftex,bookmarks,breaklinks,bookmarksopen,bookmarksnumbered,colorlinks,linkcolor=LINKCOL,linktocpage,citecolor=CITECOL,urlcolor=URLCOL,pdfpagemode=UseOutline]{hyperref}

\definecolor{TITLECOL}{rgb}{0.1,0.2,0.7} 
\definecolor{SECOL}{rgb}{0.1,0.2,0.7} 
\definecolor{CONTENTSCOL}{rgb}{0.1,0.2,0.7} 
\definecolor{SSECOL}{rgb}{0.25,0,0.48} 
\definecolor{SSSECOL}{rgb}{0.2,0.08,0.53} 
\definecolor{FINCOL}{rgb}{0.01,0.3,0.07}

\def\coloredtitle#1{\title{\textcolor{TITLECOL}{#1}}} 
\def\coloredauthor#1{\author{\textcolor{CITECOL}{#1}}} 

\definecolor{URLCOL}{rgb}{0,0.17,0.43} 
\definecolor{LINKCOL}{rgb}{0.05,0.4,0} 
\definecolor{CITECOL}{rgb}{0.35,0,0.48} 

\def\sec#1{\section{\textcolor{SECOL}{#1}}}
\def\ssec#1{\subsection{\textcolor{SSECOL}{#1}}}
\def\sssec#1{\subsubsection{\textcolor{SSSECOL}{#1}}}
\def\usec#1{\section*{\textcolor{SECOL}{#1}}}

\definecolor{URLCOL}{rgb}{0.1,0.2,0.7} 
\definecolor{LINKCOL}{rgb}{0.1,0.2,0.7} 
\definecolor{CITECOL}{rgb}{0.1,0.2,0.7} 
\def\coloredauthor#1{\author{\textcolor{PREPRINTCOL}{#1}}} 
\def\sec#1{\section{\textcolor{PREPRINTCOL}{#1}}}
\def\ssec#1{\subsection{\textcolor{PREPRINTCOL}{#1}}}
\def\sssec#1{\subsubsection{\textcolor{PREPRINTCOL}{#1}}}
\def\usec#1{\section*{\textcolor{PREPRINTCOL}{#1}}}

\definecolor{lightgray}{gray}{0.8}

\def\F{_{\sss F}}

\def\bea{\begin{eqnarray}}
\def\eea{\end{eqnarray}}
\def\ben{\begin{equation}}

\def\een{\end{equation}}
\def\benu{\begin{enumerate}}
\def\enu{\end{enumerate}}
\def\bei{\begin{itemize}}
\def\eei{\end{itemize}}

\def\br{{\bf r}}
\def\bR{{\bf R}}

\def\n{n}

\def\sss{\scriptscriptstyle\rm}

\def\g{_\gamma}
\def\l{^\lambda}

\def\x{_{\sss X}}
\def\c{_{\sss C}}
\def\s{_{\sss S}}
\def\xc{_{\sss XC}}

\def\F{_{\sss F}}

\def\LDA{^{\rm LDA}}

\def\GEA{^{\rm GEA}}

\def\PBE{^{\rm PBE}}
\def\PBERPA{^{\rm PBE-RPA}}

\def\TF{^{\rm TF}}

\def\unif{^{\rm unif}}
\def\ee{_{\rm ee}}

\def\ac{_{\sss AC}}
\def\ax{_{\sss AX}}
\def\half{\frac{1}{2}}

\def\RPA{^{\rm RPA}}
\def\z{_\zeta}
\def\nh{n_{\sss HOMO}}
\def\smooth{^{\rm smooth}}
\def\nob{^{\rm nob}}
\def\AE{^{\sss AE}}

\usepackage[normalem]{ulem}
\definecolor{darkgreen}{rgb}{0.0,0.66,0.0} 
\definecolor{darkred}{rgb}{0.75,0.00,0.0}

\begin{document}
\coloredtitle{
Locality of correlation in density functional theory}

\coloredauthor{Kieron Burke}
\affiliation{Department of Chemistry, 
University of California, Irvine, CA 92697}

\coloredauthor{Antonio Cancio}
\affiliation{Department of Physics and Astronomy, Ball State University,
Muncie, IN 47306}

\coloredauthor{Tim Gould}\affiliation{Qld Micro- and Nanotechnology Centre, %
Griffith University, Nathan, Qld 4111, Australia}

\coloredauthor{Stefano Pittalis}
\affiliation{CNR-Istituto di Nanoscienze, Via Campi 213A, I-41125 Modena, Italy}

\date{\today}

\begin{abstract}
The Hohenberg-Kohn density functional was long ago shown to
reduce to the Thomas-Fermi approximation in the non-relativistic semiclassical
(or large-$Z$) limit for all matter, i.e, the kinetic energy becomes
local.  Exchange also becomes local in this
limit.  Numerical data on the correlation energy of atoms
supports the conjecture that this is also true for correlation, but much
less relevant to atoms.  We illustrate how expansions around
large particle number are
equivalent to local density approximations and their strong relevance to
density functional approximations.  
Analyzing highly accurate atomic correlation energies, we show that
$E\c \to -A\c\, Z \ln Z +  B\c Z$ as $Z \to \infty$, where $Z$ is
the atomic number, $A\c$ is known, and we estimate
$B\c$ to be about 37 millihartrees.  
The local density approximation yields $A\c$ exactly, but a very incorrect
value for $B\c$, showing that the local approximation is
less relevant for correlation alone.
This limit is a benchmark for the non-empirical construction of
density functional approximations.  We conjecture that, beyond atoms,
the leading correction to the local density approximation
in the large-$Z$ limit generally takes this form, but with $B\c$
a functional of the TF density for the system.
The implications for construction of approximate density functionals are
discussed.
\end{abstract}


\maketitle


\sec{Introduction}

Kohn-Sham density functional theory\cite{KS65} enjoys remarkable popularity, being used in more
than 30,000 papers last year.\cite{PGB15}  Essentially all these calculations use some approximation
to the exchange-correlation energy, $E\xc$, as a functional of the (spin)-densities.\cite{FNM03}
Although many hundreds of such approximations exist and appear in standard codes, most
calculations are run with one of a few approximations.  These standard approximations
\cite{B88,LYP88,Bb93,PCVJ92,PBE96}
have been around for almost twenty years and their successes and failures are well-known.
Of course there are many excellent improvements beyond these approximations for
specific purposes.\cite{KK08,DRSL04}

Although the exact theory of DFT is well-established\cite{L79,L81} and rarely questioned these days,
it is sometimes claimed that there is no systematic approach to the overall construction of 
approximate functionals.  In fact, this is not true.  More than 40 years ago, Lieb and Simon
proved that Thomas-Fermi theory become relatively exact for the total (non-relativistic)
energy of any system in a limit in which both the particle number $N$ beomes large and
the coordinates are scaled.  For atoms,
this amounts to $Z\to\infty$, where $Z$ is the
nuclear charge, while keeping the atom neutral ($Z=N$).
But the limit also applies to all molecules and solids,
once their bond lengths are simultaneously
scaled.\cite{LS77}
This limit then implies a systematic
approach to the construction of density functionals, by finding density functional corrections
to the TF approximation that yield the exact leading corrections to the TF energy.
Unfortunately, even in very simple situations, deriving and using these leading corrections
can be very demanding, and they will not in general
be explicit density functionals.\cite{ELCB08,ECPG15,RLCE15}
Nonetheless, such functionals are in principle well-defined, if not easily approximated.

The modern world uses the KS variant of DFT, in which only the XC energy
needs approximating.  Schwinger demonstrated\cite{S80,S81} that, for atoms, the local density
approximation for exchange becomes relatively exact in the
large-$Z$ limit, and a proof (for non-singular potentials) was given by Conlon in the
general case.\cite{C83}  A rigorous proof for atoms was given by Fefferman and Seco.\cite{FSb94}
The next correction was estimated by Elliott and Burke\cite{EB09} and shown to be accurately
reproduced by both the B88 functional\cite{B88} and the exchange part of 
PBE.\cite{PBE96}  Thus, at the GGA
level, the most commonly-used approximations (at least 70\% of present calculations\cite{PGB15})
reproduce the leading correction to LDA in this limit.  
This insight was used (in an inverted piece of logic) to
construct the exchange functional in PBEsol,\cite{PRCV08} which is useful for calculating the
equilibrium properties of solids (but not their thermochemistry).  The tension between 
accurate energetics and bond lengths\cite{ZY98,PBE98} suggests that the GGA form is
insufficient to capture all aspects of the leading correction to LDA accurately.\cite{PRCV08}
The asymptotic expansion for exchange has been used to construct new
functional approximations.\cite{CTDC16,LFCD11,CFLD11}

The present paper explores the next logical step in this analysis:  Does the correlation
energy alone also become relatively exact in LDA in this limit?  If so, what is the leading correction,
and do standard approximations capture this?   
These are very fundamental questions about the nature of the XC functional, and our ability
to approximate it with semilocal functionals.  Such questions concern properties of the
exact functional that are highly relevant to approximations.

We cannot hope to answer this question in general at present, but we can try to answer this
question in the one case where we have sufficient data: atoms.
By very carefully analyzing accurate correlation
energies\cite{MT11} for a sequence of non-relativistic atoms up to $Z=86$, we find 
they can be fitted to the following form:
\ben
E\c \to -A\c\, Z \ln Z +  B\c\, Z. 
\label{eqnone}
\een
The first constant, $A\c$, is about 20.7 mHa, and is derived from the
uniform electron gas 
and confirmed analytically in Ref.~\onlinecite{KR10}.
The second constant, $B\c$, is a property of all atoms and is highly {\em inaccurate}
in LDA.  From theory and our calculations, we
estimate $B\c$ to be
37.2 mHa, whereas in LDA it is found analytically to be -4.5 mHa.
However, even this very limited information yields much insight into the nature of 
the correlation energy functional.  At a very elementary level, we see why it has
been so much more difficult to identify than the exchange form: The $1/|\br-\br'|$
behavior of the Coulomb potential produces a $Z \ln Z$ term,
so that only at astronomical values of $Z$ does the LDA correlation become relatively
exact.  
The fact that $B\c\LDA$ is negative helps explain
the huge overestimate of LDA correlation energies.  We also note that it is {\em only}
the high-density limit of LDA that is relevant to atoms, not the entire density
dependence of the uniform gas.  Thus, within this analysis, the LDA is only particularly
relevant for Coulomb matter in the extreme high-density limit.  Of course, for a simple valence
metal solid which is well-approximated by a uniform gas, the LDA is needed to produce an
accurate correlation energy.  Moreover, for energy differences and derivatives with
respect to nuclear coordinates that are relevant to bond lengths, more terms in the
LDA correlation energy might be relevant.

A first modern attempt to extract the behavior of the correlation energies for large non-relativistic atoms
was made by Kunz and Rueedi.\cite{KR10} They complemented analytical work based on
an RPA-like approximation 
with numerical values from Clementi and Corongiu~\cite{Clementi1997} up to 
$Z \!=\!  54$. 
Their estimate for $B\c$ is inaccurate, due to inaccuracies in the data and an insufficiently
precise extrapolation.\cite{footnote1}

The layout of this article is a little unusual.  Almost half the material is introductory,
explaining how the Lieb-Simon limit is relevant to approximating density functionals for
any component of the energy: Kinetic, exchange, or correlation.  This is needed to 
understand the relevance of the results for correlation.  Part of doing this is
a simple illustration of how the large-$N$ limit of a series
can be used to accurately approximate its value for {\em all} values of $N$, including
even $N=1$ (or less).   We also use this to give our definition of non-empirical
parameters.   We next give a substantial amount of background material which also serves
to define our notation.  The main body of the paper is the careful 
extrapolation, from the results we have for finite $Z$, to the large-$Z$ 
limit.  We test these extrapolations on
cases where we know the answer, and show the results are independent of the details of our
procedure.
In the final section, we discuss the relevance of our
results for density functional approximations.

We use atomic units throughout, so all energies are in Hartrees and all distances in Bohr radii.
All calculations employ spin DFT and are entirely non-relativistic.

We beg the forbearance of the reader, as we close with
a simple mathematical illustration of
the nature and usefulness of expansions as $N\to\infty$, which is
the type of expansion we will apply to density functionals.
Later, we will tie this particular
example to DFT, but for now, consider only the mathematics.

\begin{figure}[htbp]
\includegraphics[width=\columnwidth]{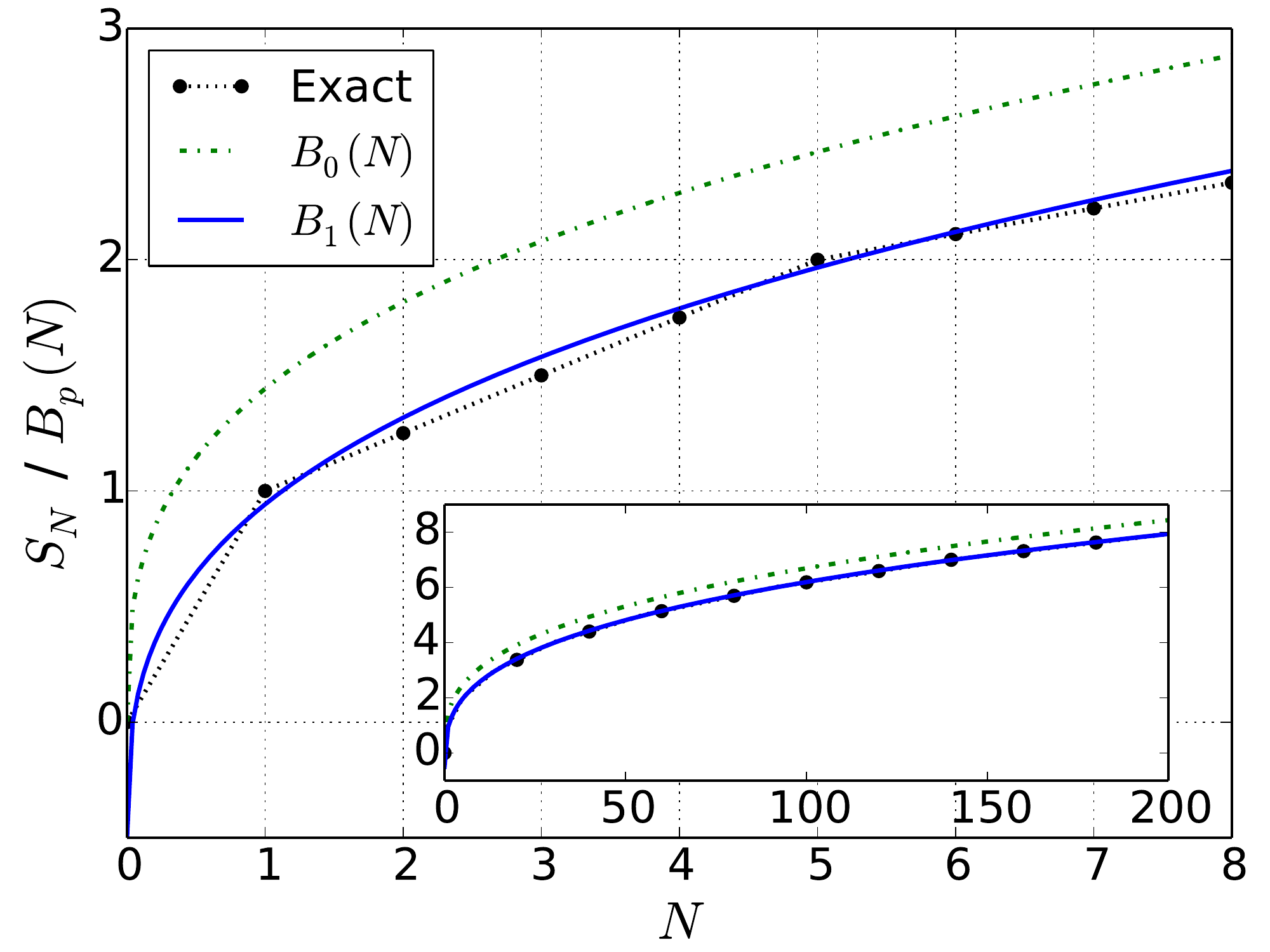}
\caption{Partial sum $S_N$ as a function of $N$,
and its large-$N$ asymptotic expansion: including just the 
leading term in $B_0(N)$, and adding the first correction in $B_1(N)$.
Inset shows
the same plots, but for significantly larger $N$.}
\label{sA}
\end{figure}
Consider an infinite series with terms:
\ben
1,1/4,1/4,1/4,1/4,1/9,\cdots,1/9,1/16,\cdots,1/16,1/25,\cdots
\een
i.e., a list of $1/n^2$, but where each value is repeated $n^2$ times.
If $f_n$ is the $n$-th term in this series, suppose we wish to
find the partial sum:
\ben
S_N= \sum_{n=1}^N\, f_n.
\een
This function is plotted in Fig. \ref{sA}.

Clearly the sum is unbounded as $N\to\infty$.  To expand the series for large $N$,
consider those values at which $S$ is an integer, a condition analogous
to filling an atomic shell:
\ben
S_{M(j)}=j,
\label{eqa}
\een
where 
\ben
M(j)=\sum_{k=1}^j k^2 = j(j+1)(2j+1)/6.
\label{eqb}
\een
Expanding $j(M)$ in inverse powers of $M^{-1/3}$, we find
\ben
j(M) = (3 M)^{1/3} - 1/2-\dots
\label{eqc}
\een
yielding
\ben
S_N \to b_0(N) + b_1 (N) + b_2(N) +\dots,
\label{eqd}
\een
where $b_0(N)= (3N)^{1/3}$, $b_1(N)=-1/2$, etc.  We then define the
sum up to a given order in $N^{1/3}$:
\ben
B_p(N) = \sum_{k=0}^p b_k(N)
\label{Bp}
\een
and compare these with the exact curve in
Fig.~\ref{sA}.
The most important point for now is that, including only the first two terms,
yields a very reasonable approximation to $S_N$, i.e., the large-$N$ expansion
can be used to approximate the series for {\em any} value of $N$, all the
way down to $N=1$.
Thus, even though we might only care about, e.g., $N < 100$, we can
find good approximations by studying the approach to the large-$N$ limit.
We will return to this example several times in this article.  For now,
we also point out several features:
\bei

\item The leading term, $B_0(N)=(3N)^{1/3}$, is always an overestimate,
being 44\% too high at $N=1$, with the error decreasing with increasing $N$.

\item Inclusion of the second term reduces the error to a 6\% underestimate
for $N=1$, and now the sign of the error varies.  This is almost always a
better approximation than $B_0(N)$, except near $N=0$, where $B_0$ gives
the exact answer, while $B_1=-1/2$.
A physical realization of this expansion and interpretation of
$B_0$ and $B_1$ is discussed in Sec.~\ref{hydrogenic}.



\item The first two terms are smooth functions of $N$, but the exact partial sum $S_N$ has
kinks wherever $S_N$ passes through an integer.

\item 
The next term in the series, $b_2(N)$, behaves as $N^{-1/3}$.
It is {\em not} smooth but instead 
periodic across a shell; thus it (and
all subsequent terms) require a more 
careful analysis than that for filled shells.\cite{E88}  
We discuss $b_2$ further in Sec.~\ref{IIIF}.
\eei


\sec{Background and notation}

\ssec{Kohn-Sham DFT}
The formal basis of DFT was established first by Hohenberg and Kohn,\cite{HK64}
but we use the more general formulation of Levy here.\cite{L79,L81}  The ground-state
energy of any electronic system can be found from an exact variational
principle in the density:
\ben
E=\min_n \left\{F[\n] + \int d^3 r\, \n(\br)\, v(\br) \right\}
\een
where $v(\br)$ is the one-body potential,  the minimization is over
all normalized non-negative densities with finite kinetic energy,\cite{L81} and
\ben
F[\n]=\min_{\Psi\to\n} \langle \Psi |\, \hat T + \hat V\ee \,| \Psi \rangle,
\een
where $\hat T$ is the kinetic energy operator, $\hat V\ee$  is the electron-electron
repulsion, and 
the minimization is over all normalized antisymmetric wavefunctions.

The original DFT was that of Thomas-Fermi,\cite{T27,F28} in which the kinetic
energy of the electrons is approximated with a local approximation:
\ben
T\TF[\n] = A\s\int d^3r\, \n^{5/3}(\br)
\label{TTF}
\een
where $A\s= 3 (3\pi^2)^{2/3}/10$, 
and their interaction energy is approximated by the Hartree electrostatic
self-energy of the density:
\ben
V\ee\TF[\n] = U[\n] = \half \int d^3r\int d^3r'\, \frac{\n(\br)\, \n(\br')}{|\br-\br'|}
\een
Applying such approximations to the total energy, and minimizing to find the
density of a system, leads to self-consistent TF theory, which was used for several
decades in materials calculations.\cite{Sa91}  The TF equation for an atom is now 
iconic,\cite{T27,F28,LCPB09} and was
solved numerically in the original papers.  Energies are typically accurate to within
about 10\%, but the theory is too crude for modern chemical and materials purposes.\cite{Sa91}

The variant in use in almost all modern electronic structure calculations
is Kohn-Sham (KS) DFT, which posits a fictitious set of non-interacting
electrons with the same ground-state density.\cite{KS65}  These electrons satisfy
the Pauli principle and obey the KS equations:
\ben
\left\{-\half \nabla^2 + v\s(\br) \right\} \phi_i(\br) = \epsilon_i\,
\phi_i(\br).
\een
Writing the energy in terms of these orbitals, 
\ben
E = T\s + U + V + E\xc,
\label{Es}
\een
where $T\s$ is the kinetic energy of the KS orbitals, $U$ their Hartree
energy, and $E\xc$ is {\em defined} by Eq. (\ref{Es}).
For simplicity, we write all equations here in terms of DFT, but in 
practice, and in our calculations, we use spin-DFT.\cite{BH72}

Kohn and Sham also wrote down the most primitive approximation to use
in their equations,\cite{KS65} the local density approximation (LDA) for XC:
\ben
E\xc\LDA = \int d^3r\, e\unif\xc(\n(\br))
\een
where $e\xc\unif(\n)$ is the XC energy density of a uniform gas of density $\n$,
which is now known very accurately.\cite{CA80,VWN80,PW92}  This was used for a generation in
solid-state physics,\cite{JG89,J15} but almost always significantly overbinds molecules, so
it never became widespread in chemistry.  However, most of the properties
of molecules in LDA calculations (such as bond lengths and vibrational frequencies)
are surprisingly accurate.\cite{JG89}

At first, it was thought that LDA might be improved using the
gradient expansion,\cite{DG90} which includes the leading corrections to the LDA energy
for a sufficiently slowly
varying gas:
\ben
E\xc\GEA = E\xc\LDA + \int d^3r\, g\xc(\n)\ s^2(\br)
\een
where $s=|\nabla\n|/(2 k\F n)$, $k\F$ is the local Fermi wavevector,
and $g\xc(\n)$ has been derived from many-body theory,\cite{KL88,MB68}
at least in the high-density limit.  However, in many cases, these corrections
worsened LDA results for atoms and molecules.  This lead ultimately to the
development of modern {\em generalized} gradient approximations,\cite{LP77} which 
include more general functions of $s$ (but not higher gradients), and led
to the widespread use of KS-DFT in chemistry and materials today.\cite{B14}
In this language, we can then understand that $T\TF=T\s\LDA$, i.e., the Thomas-Fermi
theory approximates the kinetic energy as a local functional of the density for non-interacting
electrons.

The standard way to think about XC is in terms of the strength of the
Coulomb interaction.\cite{FNM03}  In DFT, we put a coupling constant $\lambda$
in front of $V\ee$ and imagine varying it, keeping the density fixed.
Then, for finite systems,
\ben
E\xc\l = \lambda E\x + E\c\l
\een
where the correlation energy contains only $\lambda^2$ and higher powers of 
$\lambda$.  Truncation at first order yields exact exchange (EXX), 
the DFT version
of Hartree-Fock,\cite{ED11} (HF) which yields ground-state energies almost identical to those of
HF.\cite{GE95}  

One can relate this $\lambda$-dependence to
coordinate scaling of the density\cite{LP85}:
\ben
\n_\gamma(\br) = \gamma^3\, \n(\gamma\br),~~~~~0 < \gamma < \infty
\label{ngam}
\een
via
\ben
E\xc\l[\n] = \lambda^2\, E\xc[\n_{1/\lambda}].
\een
This can be used to write $E\xc$ entirely in terms of a purely potential
contribution, the 
adiabatic connection formula.\cite{HJ74,LP75,GL76}
Furthermore, in the absence of degeneracies,
\ben
T\s[\n]=\lim_{\gamma\to\infty}F[\n\g]/\gamma^2,
\label{Tsg}
\een
and
\ben
E\x[\n]=
\lim_{\gamma\to\infty}E\xc[\n\g]/\gamma.
\label{Exg}
\een


\ssec{Lieb-Simon $\zeta$-scaling}

We begin with a definition of scaling to large $Z$.  
For any system of $N$ non-relativistic electrons
with one-body potential $v(\br)$, we define a $\zeta$-scaled system
following Lieb and Simon:\cite{LS77}
\ben
v_\zeta(\br)=\zeta^{4/3}\, v(\zeta^{1/3}\br),~~~~~~~~N_\zeta=\zeta\, N,
\een
where $\zeta$ is a positive real number.
As $\zeta\to\infty$, this corresponds to simultaneously scaling the coordinates
and increasing the particle number.  For atoms or ions,
\ben
v(\br)=-Z/r,~~~~v_\zeta(\br) = -\zeta Z/r,
\een
i.e., $\zeta$-scaling is simply  the same as changing the number of protons
and electrons by the same fraction.
For molecules and solids, all nuclear separations $\bR$ also scale as $\bR/\zeta^{1/3}$.
The complementary density scaling is
\ben
\n_\zeta(\br) = \zeta^2\, \n(\zeta^{1/3}\br).
\een
This differs from the usual coordinate scaling of Eq. (\ref{ngam}) as the particle number also
changes. Analogous to that case, the $\zeta$-scaled density is, in general, {\em not}
the ground-state density of the $\zeta$-scaled potential, except within the TF approximation.

\ssec{Lieb-Simon theorem}

Lieb and Simon\cite{LS73} (LS) 
rigorously proved that, as $\zeta\to\infty$,
\ben
\lim_{\zeta\to\infty} \frac{E(\zeta)-E\TF(\zeta)}{E(\zeta)} \to 0
\een
for all non-relativistic Coulombic systems.
For example, the percentage error of the energy
in a TF calculation for an atom vanishes for large $Z$.
Moreover, any well-behaved integral over the density of the form\cite{L81}
\ben
I = \int d^3r\, f(\n(\br),\br)
\een
will also become relatively exact.  Although $\n\TF(\br)$ has many
well-known deficiencies (divergence at the nucleus, non-exponential decay at
large distances, missing quantum oscillations),
integrals such as $I$ over $\n^{TF}(\br)$ have vanishing relative error in this limit.


Without taking too large a leap, we assume the LS result is also true for 
non-interacting electrons in any potential,\cite{FLS15} yielding
\ben
\frac{T\s(\zeta)-T\TF\s(\zeta)}{T\s(\zeta)} \to 0,~~~~~\zeta\to\infty
\label{LSth}
\een
again, a universal limit of all systems. Thus any approximation that fails to
satisfy this limit is unlikely to be a useful starting point for higher 
accuracy approximations for a large variety of systems.

\ssec{Illustration: Hydrogenic atoms \label{hydrogenic}}

The simplest useful example of $T\s$ is to consider non-interacting electrons
in a hydrogen potential, $-1/r$.   Their energies are $-1/2n^2$, where $n$ is the principal
quantum number.  Via the virial theorem, $T\s=-E$, and accounting for double
occupation of the orbitals, in fact
\ben
T\s(N) =\half\, S_{N/2}
\een
where $S_N$ is the partial sum from the introduction.  Thus our purely mathematical
example is in fact a relevant example.  The TF density for this problem is simply:\cite{HL95}
\ben
\n\TF(r)=\frac{4Z}{\pi^2 r^3_c}\, \left(\frac{r_c}{r}-1\right)^{3/2} \Theta (r_c-r),
\label{nTFbohr}
\een
where $r_c \!=\!  (18/Z)^{1/3}$, and $\Theta$ is the Heaviside step function.
This situation is similar to the interacting
case, in that the TF density is singular at the origin, missing oscillations, and does not
decay correctly at large $r$. 
Insertion of Eq. (\ref{nTFbohr})
into Eq. (\ref{TTF}) yields precisely $(3N/2)^{1/3}/2$, the leading
term $B_0(N/2)/2$ in the expansion for large $N$.  This is consistent with 
Eq. (\ref{LSth}).  Thus, this reasoning says that, for an orbital-free theory to
take advantage of the LS result, it should recover both this value
and the next correction\cite{S52} (-1/2) in the large-$Z$ limit.  Of course, this is unlikely
to be sufficient to determine a usefully accurate orbital-free DFT.  But if
the reasoning commonly applied to XC is applied to this problem, i.e., that
approximate functionals should build in those conditions that they
can satisfy, it is a {\em necessary} exact condition.
No approximate kinetic energy functional known at present produces the correct value for the coefficients of this expansion
for all electronic systems.\cite{LCPB09}

\ssec{Exchange-correlation analog}

This section addresses a fundamental question about DFT in chemistry
and materials science, which is: How can such simple approximations, such as LDA,
GGA or even hybrids, possibly yield usefully accurate results for such a demanding many-body
problem?   While their failures for many specific properties and systems are
myriad, their successes are legion.  How can this be, since everyone knows
how demanding and complicated the many-body problem is?

The seeds of the answer are in the Lieb-Simon theorem.  As $Z$ grows, local approximations
become more accurate for the energy and other integrated quantities (but not pointwise
for densities, energy-densities, or potentials).  The many-body
problem becomes simple in two extremes: $N=1$ and $N\to\infty$.  All the detailed
behavior of the wavefunctions is averaged away, as quantum oscillations become more
and more rapid in space, having less and less effect on system averages.  

This then provides a systematic approach to constructing DFT approximations, but one which
is entirely different from the traditional approach in terms of many-body theory and expansions
in $\lambda$, the electron-electron interaction.
The Thomas-Fermi approximation becomes relatively exact for large $N$.  
Note that we do not claim that our systems of interest have large $N$, and indeed,
a direct large-$N$ expansion of the energy is often not sufficiently accurate.  Instead,
we apply the large-$N$ DFT approximation self-consistently to the problem at hand.
We are using the density functional that is exact at large $N$, which is much
more accurate than simply taking the large-$N$ behavior of our specific system.
In terms of our hydrogenic illustration, a GGA for $T\s$ can be constructed that
automatically recovers the first two terms in the expansion of the energy, but
that is more accurate for small $N$ than the asymptotic expansion alone.\cite{LCPB09}

We now conjecture that the following statement is true for KS-DFT:
\ben
\lim_{\zeta\to\infty} \frac{E\xc(\zeta)-E\xc\LDA(\zeta)}{E\xc(\zeta)} \to 0.
\een
We call this a conjecture, because the Lieb-Simon theorem for TF theory
has been proven rigorously in a mathematical physics sense,\cite{LS83} but no
such general demonstration yet exists for $E\xc$.  However, all evidence
suggests that this conjecture is correct.  For example, few would doubt that,
in a limit in which the number of particles is growing, $E\c/E\x\to 0$, so that
the statement need only be proven for $E\x$.
This was demonstrated very carefully by Schwinger\cite{S80,S81,ES82,ES84,ESb84,ESc84,
ES85,ESb85,ESc85}
for atoms, and detailed in the pioneering
book of Englert.\cite{E88}  It was later proven rigorously for 
atoms.\cite{FSb94}
It can be proven for all
systems if the singularity at the nuclei is smoothed,\cite{C83} but no general mathematical
proof has been
given for Coulomb interactions.  For our purposes, this is sufficient.

Virtually all modern XC approximations reduce to $E\x\LDA$ for
uniform densities, which means they also do so for inhomogeneous systems as $\zeta \to \infty$.
Assuming the conjecture is correct, this makes sense as $E\x\LDA$ is then 
a {\em universal} limit for all systems, and its relevance has nothing to do
with the rapidity of the density variation (which is never slow for realistic
finite systems).  By studying the
exchange energies of noble gas atoms of increasing $Z$, one can\cite{EB09} deduce $E\x\LDA[\n]$,
i.e., it is a simple limit of {\em all} systems, one in  which the many-body
problem becomes very difficult (the number of particles is diverging), but in which
the density functional for the integrated quantity simplifies enormously.  

As $\zeta\to\infty$, the coupling between electrons becomes weak, just as 
it happens for
$\gamma\to\infty$, in coordinate scaling.  Thus this limit is weakly correlated.  But
the number of particles is also growing, and the number of pair-interactions is growing
even faster.  In this limit,\cite{PCSB06}
\ben
V\ee[\n\z] \to U[\n] + \zeta^{5/3}\, E\x\LDA[\n]+\dots
\een
If we compare this with the usual coordinate scaling, and speak in
terms of the coupling constant, as $\zeta$ grows, the effective coupling
constant is shrinking.  But due to the increased number of interactions,
the net effect is as given above.
The analog to Eq. (\ref{Tsg}) using the Lieb-Simon theorem
is
\ben
T\s\TF[\n] + U[\n]
=\lim_{\zeta\to\infty} F[\n_\zeta]/\zeta^{7/3}
\een
while for XC, the analog to Eq. (\ref{Exg}) is
\ben
E\x\LDA[\n]
=\lim_{\zeta\to\infty} E\xc[\n_\zeta]/\zeta^{5/3}.
\een

We can then argue that, if standard DFT approximations work reasonably
well for weakly-correlated systems, it is because the density-dependence of the XC energy,
found self-consistently in the KS scheme, is well-approximated by the limiting density
functional.  Most molecular systems at or near equilibrium bond lengths are
weakly correlated, so LDA yields usefully accurate values in this region.  While LDA
is insufficiently accurate for the energy difference between an equilibrium
bond and dissociated atoms, its error is extremely systematic,
and the next correction to the energy in the form of a GGA works reasonably well.

This is entirely analogous to the starting approximation for most many-body treatments.
Hartree-Fock becomes relatively exact as the electron-electron repulsion becomes weak.
But in practice we apply it directly to any many-electron system with the full electron-electron
repulsion, and find usefully accurate descriptions (although not bond-energetics) from the
self-consistent solution of those equations.
LDA becomes relatively exact in a different limit, as discussed
above.
In this sense, LDA is as non-empirical 
as HF, and is substantially more accurate for bond energies.

\ssec{Locality principle}

Here we claim that a 
{\em locality} principle is at work in DFT:  Many of the successes and failures 
of DFT approximations to XC can be understood in terms of the expansion of the 
functional around the large-$Z$, i.e., local limit. 
The success of LDA for real materials (not slowly varying gases)
has nothing to do with the uniform gas per se.  It is a universal
limit of {\em all} quantum systems, and that limit is most easily calculated from 
the uniform gas.  
The real question is:  For real molecules and materials, does 
LDA dominate, and if so, how large are the corrections?
LDA works as well (or as badly) as it does for weakly correlated systems because,
for those systems, the density-dependence of the
XC energy functional is moderately accurately approximated
by its limiting form (i.e., a local one).  For such systems, and only such systems,
inclusion of the next correction should
improve its accuracy.  
Thus most molecules and many materials at equilibrium are weakly correlated.
LDA works reasonably well, and 
standard GGA's usually improve energetics substantially.  On the other
hand, static correlation is a signal that this expansion is failing, and hybrid
functionals\cite{Bb93,BEP97} are a crude attempt to account for this.

\def\bl{^{\rm{BL}}}
We therefore define the beyond-local (BL) XC energy functional as
\ben
E\xc\bl[\n] = E\xc[\n]-E\xc\LDA[\n]
\een
and an obvious question arises:  If LDA becomes relatively exact for $E\xc$
at large $\zeta$, what approximation for $E\bl\xc[\n]$
becomes relatively exact in this limit?  This is the natural expansion
of which LDA is the leading order, and one can hope that accurate inclusion
of the next order will lead to accurate energetics for weakly correlated systems.
Presumably, standard GGA's are crude examples of such corrections.

\ssec{A simple example: Exchange}

The answers for exchange are simple, and well-established.\cite{EB09}  In this case, for atoms,
\ben
E\x \to -A\x\, Z^{5/3} + B\x\, Z + \dots
\een
The dominant term is exact within LDA, given by
\ben
E\x\LDA [\n] = - \frac{3}{4}\left(\frac{3}{\pi}\right)^{1/3}\int d^3r\, \n^{4/3}(\br)
\label{ExLDA}
\een
To evaluate this, we write the TF atomic density in dimensionless form\cite{E88}:
\ben
\n\TF(r) = \frac{Z^2}{4\pi a^3} f(x)
\label{nTF}
\een
where $x=Z^{1/3}r/a$ and $a=(3\pi/4)^{2/3}/2$.  Inserting this into Eq. (\ref{ExLDA})
yields\cite{LCPB09}
\ben
A\x = \left(\frac{9}{2\pi^4}\right)^{1/3} M,~~~~~~~~M=\int_0^\infty dx\ x^2 f^{4/3}(x)
\een
where $M$ is
known numerically to be about 0.615434679.\cite{LCPB09}  Thus $A\x \approx 0.2208274$.
The precise value of $B\x$ has not been derived, but 
estimates for $B\x$ have been extracted numerically by careful extrapolation to the
large $Z$ limit of atoms.  $B\x\LDA \approx 0$, but $B\x \approx 0.224 $, i.e., there is a large
beyond-local contribution.  The two most commonly used\cite{B88,PBE96} generalized
gradient approximations for $E\x$ recover this beyond-local contribution rather accurately, whereas
the gradient expansion for the slowly-varying gas, when applied as an approximation to 
atoms, is smaller by about a factor of 2.\cite{PCSB06,PRCV08}  

\ssec{Non-empirical constants \label{nonemp}}

At this point, we make a small aside about what we mean by the term non-empirical.  For simplicity,
return to our mathematical example.   Suppose we argue (or notice from examples) that
the leading term must grow as $N^{1/3}$, and we use some set of tabulated values of $S_N$
to fit the coefficient in $B_0(N)$.  If we used just the point $N=1$, we would get this
coefficient quite wrong (1 instead of $3^{1/3}\approx 1.4422$).  
If we use the first 10 points, we get a more accurate answer.
But if, instead, we use tabulated values of $S_N$ to
extrapolate the $N\to\infty$ value, we get our best estimate of the
non-empirical value.
For example, knowing $S_N$ for $N=1$ to $100$,
and the form $A N^{1/3} + B$, we can estimate
$A=1.4458$ (versus the exact value of $3^{1/3}=1.4422$) 
and $B=0.525$ (versus $1/2=0.5$), by using only the last 50 values.  

By contrast, simple fitting (such as a polynomial fit) to a
range of finite data, including small $N$ but without
accounting for the asymptotic analytic form,  will usually produce a numerically
more accurate value within the range of the data, but will fail as $N$ grows larger,
and might be disastrous when attempting to find the next correction.  Note also that it
is vital to know the correct form, to guarantee relative exactness as $N$ grows, and to allow
investigation of higher-order terms.

Thus, by using tabulated values and the correct form
of the expansion, we can estimate a non-empirical coefficient,
even though we might not have the ability or energy to derive its value analytically.
The larger the values of $N$ that we use, and the more terms in the form, the more
accurate our estimate can be.  
We denote the result as non-empirical, even though it has not been derived analytically,
and we do not get the exact analytic value, just a best estimate of it.

\sec{Atomic correlation}

\ssec{Data}
\sssec{Quantum chemical data}

\begin{figure}[htbp]
\includegraphics[width=\columnwidth]{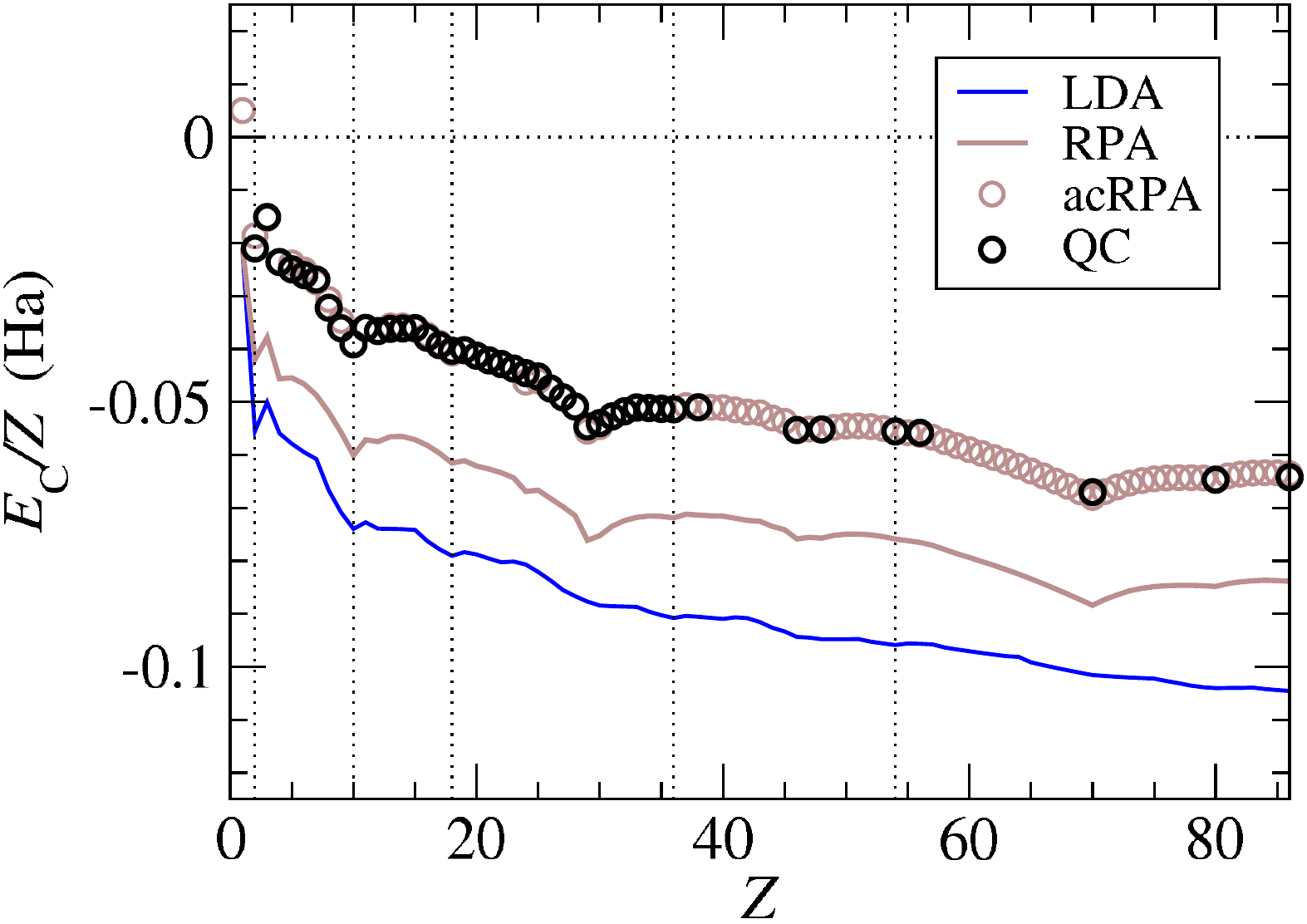}
\caption{Correlation energy per electron versus $Z$ for accurate quantum 
chemistry (QC) calculations (black circles), 
the RPA (brown solid line), 
corrected RPA data (brown circles),
and the local density approximation (LDA) of DFT.
Circle size matches the maximum reported error of QC data.
\label{Ecall}
}
\end{figure}

We are almost ready to begin our analysis of the locality of the correlation
energy.  All our data will come from non-relativistic atomic calculations.
Our starting point is the relatively recent publication of total correlation
energies of spherical atoms\cite{MT11} up to $Z=86$ and non-spherical ones\cite{MT12} up to $Z=36$.
These add to the well-known benchmark set\cite{CGDP93} up to $Z=18$, although they are not quite
as accurate.  It is the existence of these results that make the following analysis possible.
Their values appear (with many others) in Tables~\ref{tab:EcI}
and \ref{tab:EcII}, and are plotted in Fig. \ref{Ecall},
in the form of correlation energy per electron.

The behavior of correlation energies across rows of the periodic table,
 and how it changes as one goes down
a column, will play an important role in our analysis, but the data set becomes sparse
for large $Z$.  We have therefore performed pure random-phase-approximation (RPA) calculations
for all elements up to $Z=86$,
and these results are included in Tables~\ref{tab:EcI}/\ref{tab:EcII}
and Fig. \ref{Ecall}.
We find that these RPA results can be ``asymptotically corrected'' 
using a simple formula \eqref{dEcRPA} described later in the manuscript.
Using only noble gases as a fitting set, this 
asymptotically corrected RPA (acRPA) nearly matches QC trends for all
atoms for which QC data is available. We thus use the acRPA to 
`fill in' gaps in the reference data set.

\sssec{RPA calculations}
The adiabatic connection formula\cite{HJ74,LP75,GL76} combined
with the  fluctuation-dissipation theorem and RPA is fast
becoming a \emph{de facto} standard\cite{EBF12,LHGAKD10,BGKN12}
for calculations of \emph{nearly} quantum chemical accuracy energy
differences. Unfortunately, it is well known that RPA gives poor
estimates of \emph{absolute} correlation energies. As we
will show later, this erroneous contribution can, at least in the
atomic systems considered here, be cancelled by a simple fit,
Eq.~(\eqref{dEcRPA}), 
depending on the number of electrons only.
This thus renders corrected RPA 
suitable for accurate correlation energy calculations of atoms, on a par with
the most accurate quantum chemical benchmarks.

RPA energies are calculated using the ACFD correlation
energy formula
\ben
  E\c=\int_0^1 d\lambda\int d^3 r\,d^3 r'\,
  \frac{n_{2{\sss C}}\l(\br,\br')}{2|\br-\br'|}
  \label{eqn:ACFD}
\een
Here the pair-density
$n\l_{2{\sss C}}$ is found from the fluctuation dissipation theorem via
\begin{align}
  n_{2{\sss C}}\l(\br,\br')=&\int_0^{\infty}\frac{d\omega}{\pi}
  [\chi\l(\br,\br';i\omega)-\chi^0(\br,\br';i\omega)]
  \\
  \chi^0=&2\Re \sum_{i}f_i\phi_i(\br)\phi_i(\br')
  G(\br,\br';\epsilon_i-i\omega)
  \\
  \chi\l=&\chi^0
  +\lambda\chi^0\star\frac{1}{|\br-\br'|}\star\chi\l
\end{align}
where stars indicate spatial
convolutions. The non-interacting response $\chi_0$
takes as input the Kohn-Sham orbitals $\phi_i$ and energies
$\epsilon_i$ obeying $\{-\half\nabla^2 + v\s-\epsilon_i\}\phi_i=0$;
and KS Greens functions $G(\br,\br';\epsilon_i-\omega)$
obeying
$\{-\half\nabla^2 + v\s-\epsilon_i+i\omega\}
G(\br,\br';\epsilon_i-i\omega)=-\delta(\br-\br')$.
We work from spherical and spin symmetric groundstates
which make all calculations essentially one-dimensional.
All equations can thus be carried out using a large radial grid
which helps to reduce numerical errors in \eqref{eqn:ACFD}.
Errors are estimated to be well under 1.1~mHa per electron.
We use the same algorithms and code
as previous work.\cite{Gould2013-LEXX,Gould2013-Aff}

Equation \eqref{eqn:ACFD} is an indirect orbital function
which formally maps a Kohn-Sham potential $v\s$
and orbital occupation factors $f_i$
to a correlation energy. We must thus start from a reasonably accurate
potential if we are to expect accurate energies.
For this work, we perform RPA calculations using strictly spherical
Kohn-Sham potentials calculated using LEXX\cite{Gould2013-LEXX} theory
extended to $d$ and $f$ shells. 
LEXX yields a Hartree and exchange energy functional
of the same form as HF theory,
but in which all orbitals are systematically calculated in a
multiplicative spherically symmetric potential. Its associated
correlation energy is thus slightly larger in magnitude.
For most elements, the occupation
factors are assigned according to Hund's rules,
in accordance with theoretical and experimental evidence.
For the transition metals we use
the lowest energy orbital filling (in exact exchange theory) of the
$s$ and $d$ shells
(e.g. $4s^13d^{10}$ for Cu and $5s^14d^{4}$ for Nb), which allow ready
comparison with the results of McCarthy and Thakkar.\cite{MT12}
For the atoms in Row 6 we simply fill the orbitals according to
Hund's rules to avoid issues of (near-)degeneracy.

Ideally the orbitals would be calculated on the exact Kohn-Sham
potential $v_s(\br)$, but this is
available for only a very limited
number of atoms, forcing us to use an appropriate approximation
to the potential. The LEXX potential was chosen due to its
good 
asymptotic form and inclusion of static correlation in
open shell systems.
To ensure our RPA energies are appropriate we performed some
additional tests to study dependence on the potential.
Firstly, for C and F we evaluated RPA energies using both
the exact\cite{GT14} potential and the LEXX potential and found
the difference to be less than 0.3~mHa per electron. We also
compared energies calculated using the PBE potential and LEXX
potential across many species and found a maximum difference of
1.0~mHa per electron, a likely worst case scenario. We thus assume
our total error (numerical and error from the potential) to be
under 1.5~mHa per electron or 3\%, whichever is the larger.
This gives us, through RPA and the corrections
discussed later, accurate benchmarks for all non-relativistic
atoms with up to 86 electrons. 

\sssec{DFT calculations}

At the opposite end of the scale, one can perform atomic DFT calculations
of correlation energies rather simply, and go to much larger $Z$ than with correlated
wavefunction treatments.  For the purposes of our study, we include all $Z$ up to 
86, and use both LDA and PBE results.  We use the PW92 parametrization of the uniform
gas,\cite{PW92} the relevance of which will be discussed below.  We include PBE because, as 
shown below, its form mimics that of the exact functional in the large $Z$ limit, so that
the large-$Z$ behavior can be most easily extracted by comparison 
with that approximation.

\setlength\LTleft\parindent
\setlength\LTright\fill
\setlength\LTcapwidth\textwidth
\sssec{Results}
\begin{table}[h!]
\caption{Negative of atomic correlation energies per electron in Ha.
``Bench" is benchmark data from Refs.~\onlinecite{CGDP93,MT11,MT12}
and ``acRPA" is asymptotically corrected RPA [RPA + Eq.~\eqref{dEcRPA}].
Listed errors are the deviation from the ``bench'' value
(where available) or ``acRPA'' otherwise.
Numerical errors on ``bench'' are estimated to be within 
2.2\% for $Z\leq54$ and 3.5\% for larger atoms\cite{MT11,MT12} while errors 
on ``acRPA'' could be up to 5.5\% (including numerical errors and 
differences between EXX and HF energies). Errors on LDA and PBE
are estimated to be around 1~mHa per electron.}
\begin{ruledtabular}
\begin{tabular}{|l|r|rr|rr|rrr|}
 $N$ & Bench & RPA & Err & acRPA & Err & LDA & PBE & Err \\\hline
1 & 0.0 & 20.9 & 20.9 & -5.0 & -5.0 & 21.7 & 5.7 & 5.7 \\
2 & 21.0 & 42.0 & 21.0 & 18.6 & -2.4 & 55.5 & 20.5 & -0.5 \\\hline
3 & 15.1 & 37.8 & 22.7 & 15.4 & 0.3 & 50.1 & 17.0 & 1.9 \\
4 & 23.6 & 45.6 & 22.0 & 23.6 & 0.0 & 55.9 & 21.4 & -2.2 \\
5 & 25.0 & 45.5 & 20.5 & 23.8 & -1.2 & 57.9 & 23.1 & -1.9 \\
6 & 26.1 & 46.6 & 20.6 & 25.1 & -0.9 & 59.5 & 24.5 & -1.6 \\
7 & 26.9 & 48.7 & 21.8 & 27.4 & 0.5 & 60.8 & 25.5 & -1.4 \\
8 & 32.2 & 51.9 & 19.7 & 30.7 & -1.6 & 66.7 & 29.5 & -2.7 \\
9 & 36.1 & 55.8 & 19.7 & 34.6 & -1.5 & 70.9 & 32.5 & -3.6 \\
10 & 39.1 & 60.1 & 21.0 & 39.0 & -0.1 & 74.0 & 34.7 & -4.4 \\\hline
11 & 36.0 & 57.2 & 21.1 & 36.2 & 0.1 & 72.7 & 33.5 & -2.5 \\
12 & 36.6 & 57.5 & 20.9 & 36.5 & -0.1 & 73.9 & 34.1 & -2.5 \\
13 & 36.2 & 56.6 & 20.4 & 35.7 & -0.5 & 74.0 & 34.3 & -1.9 \\
14 & 36.1 & 56.5 & 20.4 & 35.7 & -0.5 & 74.0 & 34.6 & -1.5 \\
15 & 36.1 & 57.1 & 21.1 & 36.3 & 0.2 & 74.2 & 35.0 & -1.1 \\
16 & 37.9 & 58.2 & 20.3 & 37.4 & -0.5 & 76.3 & 36.6 & -1.3 \\
17 & 39.3 & 59.7 & 20.4 & 38.9 & -0.4 & 77.8 & 38.0 & -1.3 \\
18 & 40.3 & 61.5 & 21.2 & 40.8 & 0.4 & 79.1 & 39.1 & -1.2 \\\hline
19 & 40.2 & 61.1 & 20.9 & 40.4 & 0.2 & 78.4 & 38.5 & -1.7 \\
20 & 41.3 & 62.1 & 20.8 & 41.4 & 0.1 & 78.8 & 38.7 & -2.6 \\
21 & 42.2 & 62.7 & 20.5 & 42.0 & -0.2 & 79.6 & 39.3 & -2.8 \\
22 & 42.9 & 63.4 & 20.5 & 42.7 & -0.2 & 80.3 & 40.0 & -2.9 \\
23 & 43.7 & 64.3 & 20.6 & 43.7 & -0.0 & 80.1 & -- & -- \\
24 & 44.6 & 65.4 & 20.9 & 44.8 & 0.2 & 80.7 & 41.0 & -3.6 \\
25 & 45.2 & 66.7 & 21.5 & 46.0 & 0.8 & 82.1 & 41.7 & -3.5 \\
26 & 47.5 & 68.2 & 20.7 & 47.6 & 0.1 & 83.7 & 42.9 & -4.6 \\
27 & 49.1 & 69.8 & 20.6 & 49.2 & 0.0 & 85.5 & 44.5 & -4.7 \\
28 & 50.9 & 71.5 & 20.6 & 50.9 & 0.0 & 86.7 & 45.4 & -5.5 \\
29 & 54.6 & 73.3 & 18.7 & 52.7 & -1.9 & 87.7 & 46.2 & -8.4 \\
30 & 54.0 & 75.2 & 21.2 & 54.7 & 0.7 & 88.4 & 46.7 & -7.3 \\
31 & 52.7 & 73.5 & 20.7 & 52.9 & 0.2 & 88.6 & 46.8 & -5.9 \\
32 & 51.8 & 72.4 & 20.6 & 51.9 & 0.1 & 88.6 & 47.0 & -4.7 \\
33 & 51.0 & 71.8 & 20.8 & 51.3 & 0.2 & 88.7 & 47.2 & -3.8 \\
34 & 51.2 & 71.5 & 20.4 & 51.0 & -0.2 & 89.6 & 47.9 & -3.3 \\
35 & 51.3 & 71.6 & 20.3 & 51.1 & -0.2 & 90.3 & 48.5 & -2.8 \\
36 & 51.4 & 71.9 & 20.5 & 51.4 & 0.0 & 90.8 & 49.0 & -2.4 \\
\end{tabular}
\end{ruledtabular}
\label{tab:EcI}
\end{table}

\begin{table}[h!]
\caption
{Same as Table \ref{tab:EcI}, but larger $Z$.}
\label{tab:EcII}
\begin{ruledtabular}
\begin{tabular}{|l|r|rr|rr|rrr|}
$N$ & Bench & RPA & Err & acRPA & Err & LDA & PBE & Err \\\hline
37 & -- & 71.2 & 20.5 & 50.7 & -- & 90.4 & 48.6 & -2.1 \\
38 & 51.0 & 71.4 & 20.3 & 50.9 & -0.2 & 90.6 & 48.7 & -2.4 \\
39 & -- & 71.5 & 20.5 & 51.0 & -- & 90.8 & 48.9 & -2.2 \\
40 & -- & 71.6 & 20.5 & 51.1 & -- & 91.0 & 49.1 & -2.0 \\
41 & -- & 71.8 & 20.5 & 51.3 & -- & 90.7 & 49.3 & -2.0 \\
42 & -- & 72.1 & 20.5 & 51.6 & -- & 90.8 & -- & -- \\
43 & -- & 72.5 & 20.5 & 52.1 & -- & 91.6 & 49.9 & -2.2 \\
44 & -- & 73.0 & 20.4 & 52.6 & -- & 92.6 & 50.9 & -1.6 \\
45 & -- & 73.6 & 20.4 & 53.2 & -- & 93.3 & 51.5 & -1.6 \\
46 & 55.3 & 74.3 & 19.0 & 53.8 & -1.4 & 94.4 & 52.5 & -2.8 \\
47 & -- & 75.0 & 20.4 & 54.5 & -- & 94.5 & 52.6 & -1.9 \\
48 & 55.2 & 75.7 & 20.5 & 55.3 & 0.1 & 94.8 & 52.8 & -2.4 \\
49 & -- & 75.2 & 20.4 & 54.8 & -- & 94.8 & 52.8 & -2.0 \\
50 & -- & 75.0 & 20.4 & 54.6 & -- & 94.8 & 52.9 & -1.6 \\
51 & -- & 75.0 & 20.4 & 54.5 & -- & 94.8 & 53.0 & -1.6 \\
52 & -- & 75.1 & 20.4 & 54.7 & -- & 95.3 & 53.4 & -1.3 \\
53 & -- & 75.4 & 20.4 & 55.0 & -- & 95.6 & 53.7 & -1.3 \\
54 & 55.6 & 75.9 & 20.3 & 55.5 & -0.1 & 95.9 & 54.0 & -1.5 \\\hline
55 & -- & 76.2 & 20.4 & 55.8 & -- & 95.6 & 53.7 & -2.1 \\
56 & 55.8 & 76.5 & 20.7 & 56.1 & 0.3 & 95.7 & 53.7 & -2.1 \\
57 & -- & 77.1 & 20.4 & 56.7 & -- & 95.8 & 53.9 & -2.8 \\
58 & -- & 77.9 & 20.4 & 57.5 & -- & 96.4 & 54.4 & -3.1 \\
59 & -- & 78.7 & 20.4 & 58.3 & -- & 96.7 & 54.7 & -3.5 \\
60 & -- & 79.4 & 20.4 & 59.0 & -- & 97.1 & 55.0 & -4.0 \\
61 & -- & 80.1 & 20.4 & 59.7 & -- & 97.4 & 55.3 & -4.4 \\
62 & -- & 80.9 & 20.4 & 60.5 & -- & 97.7 & 55.6 & -4.9 \\
63 & -- & 81.7 & 20.4 & 61.3 & -- & 98.0 & 55.9 & -5.4 \\
64 & -- & 82.5 & 20.4 & 62.1 & -- & 98.1 & 56.0 & -6.1 \\
65 & -- & 83.4 & 20.4 & 63.0 & -- & 99.2 & 56.9 & -6.2 \\
66 & -- & 84.3 & 20.4 & 64.0 & -- & 99.7 & 57.3 & -6.6 \\
67 & -- & 85.3 & 20.4 & 64.9 & -- & 100.2 & 57.8 & -7.2 \\
68 & -- & 86.3 & 20.4 & 65.9 & -- & -- & -- & -- \\
69 & -- & 87.3 & 20.3 & 67.0 & -- & 101.1 & 58.6 & -8.4 \\
70 & 67.0 & 88.4 & 21.4 & 68.1 & 1.0 & 101.6 & 59.0 & -8.1 \\
71 & -- & 87.3 & 20.3 & 67.0 & -- & 101.7 & 59.1 & -7.9 \\
72 & -- & 86.4 & 20.3 & 66.1 & -- & 101.9 & 59.2 & -6.8 \\
73 & -- & 85.7 & 20.3 & 65.3 & -- & 102.0 & 59.4 & -5.9 \\
74 & -- & 85.2 & 20.3 & 64.9 & -- & 102.1 & 59.5 & -5.3 \\
75 & -- & 84.9 & 20.3 & 64.5 & -- & 102.2 & 59.7 & -4.9 \\
76 & -- & 84.7 & 20.3 & 64.4 & -- & 102.7 & 60.0 & -4.3 \\
77 & -- & 84.6 & 20.3 & 64.3 & -- & 103.1 & 60.4 & -3.9 \\
78 & -- & 84.6 & 20.3 & 64.3 & -- & 103.6 & 60.9 & -3.4 \\
79 & -- & 84.7 & 20.3 & 64.4 & -- & 103.9 & 61.2 & -3.2 \\
80 & 64.7 & 84.8 & 20.1 & 64.5 & -0.2 & 104.0 & 61.3 & -3.4 \\
81 & -- & 84.3 & 20.3 & 64.0 & -- & 104.0 & 61.3 & -2.7 \\
82 & -- & 84.0 & 20.3 & 63.6 & -- & 104.0 & 61.3 & -2.3 \\
83 & -- & 83.8 & 20.3 & 63.4 & -- & 103.9 & 61.3 & -2.1 \\
84 & -- & 83.7 & 20.3 & 63.4 & -- & 104.2 & 61.5 & -1.8 \\
85 & -- & 83.7 & 20.3 & 63.4 & -- & 104.4 & 61.7 & -1.7 \\
86 & 64.2 & 83.9 & 19.6 & 63.6 & -0.7 & 104.6 & 61.9 & -2.3 \\
\end{tabular}
\end{ruledtabular}
\end{table}

Tables \ref{tab:EcI} and \ref{tab:EcII} list all our results.
In Fig \ref{Ecall}, we plot correlation energies per electron as a function of $Z$, as well as
the RPA and LDA results.  The shapes are well-known and unsurprising.  The RPA significantly
overcorrelates the atoms, and the LDA is even worse.  Many quantum chemists 
question the value of LDA for such calculations, as the relevance of the 
uniform gas for such systems is far from
obvious.   
Indeed, from this figure alone it is 
unclear that the QC correlation energy per electron is 
diverging logarithmically -- the limiting behavior 
of Eq.~(\ref{eqnone}) derived from the uniform gas.  One could reasonably 
argue that it is approaching a constant for sufficiently large $Z$.
We demonstrate below that the latter is definitely \textit{not} the case.

Note that the difference between HF
energies~\cite{CD96,MT12} and Kohn-Sham exact-exchange (EXX)
energies (as computed in our work)
is smaller than other errors, so we ignore this difference here.

\ssec{Theory \label{theory}}

We now test the locality conjecture for correlation, and use it to extract
trends in Fig \ref{Ecall} and Table I.   The conjecture implies that
\ben
E\c[\n_\zeta] \to E\c\LDA[\n\TF_\zeta],~~~~\zeta\to\infty.
\een
Thus we begin with the local density
approximation, which uses the correlation energy of a uniform gas:
\ben
E\c\LDA[\n] = \int d^3r\, \n(\br)\ \epsilon\c\unif(\n(\br)),
\label{EcLDA}
\een
where $\epsilon\c\unif(\n)$ is the correlation energy per electron.
This is a non-trivial function of the density that is now well-known from many-body
theory and quantum Monte Carlo simulations of the uniform gas.\cite{CA80}  There are several
almost identical parametrizations\cite{VWN80,PW92} in common use.
Our interest is the large-$\zeta$ limit, which is dominated by high densities. 
In a landmark of electronic structure theory, Gell-Mann and Brueckner\cite{GB57}
applied the
random phase approximation (RPA) to the uniform gas to find:
\ben
\epsilon\c\unif = c_0\ \ln r_s - c_1 + \dots,~~~r\s\to 0
\label{epsunif}
\een
where $r\s=(3/(4\pi\n))^{1/3}$ is the Wigner-Seitz radius of density $\n$, $c_0=0.031091$,
and $c_1\RPA=0.07082$.  
In fact, Eq. (\ref{epsunif}) yields the exact high-density limit if
$c_1=0.04664$, 
the correction from RPA being due to second-order exchange.\cite{OMS66}
But, as is clear in Fig. \ref{Ecall},  LDA\cite{KS65} 
greatly overestimates the magnitude of the correlation energy of atoms
(factor of 2 or more).
For atoms with large $Z$, insert $\n\TF(\br)$ into Eq. (\ref{EcLDA}) to find:
\ben
E\c\LDA = - A\c\, Z\, \ln Z\, + B\c\LDA Z + \cdots,
\label{aclda}
\een
where $A\c=2 c_0/3=0.02073$.
$B\c\LDA$ is found by considering the contributions linear in $Z$ generated by
replacing Eq.~(\ref{nTF}) in the high-density limit of $E\c\LDA[\n]$. As a result,
\ben\label{eqBLDA}
B\LDA\c =  \frac{c_0}{3} \left[
\ln \left( 3 a^3\right)  -  I_2  \right]-c_1,
\een
where 
\ben
I_2 = \int_0^\infty dx\, f(x)\, \ln[f(x)],
\een
and is known numerically to be about -3.331462.\cite{LCPB09}
This yields 
\ben
B\c\LDA=-0.00451,
\een
as reported in Ref.~\onlinecite{PCSB06}.

\begin{figure}[htbp]
\includegraphics[width=\columnwidth]{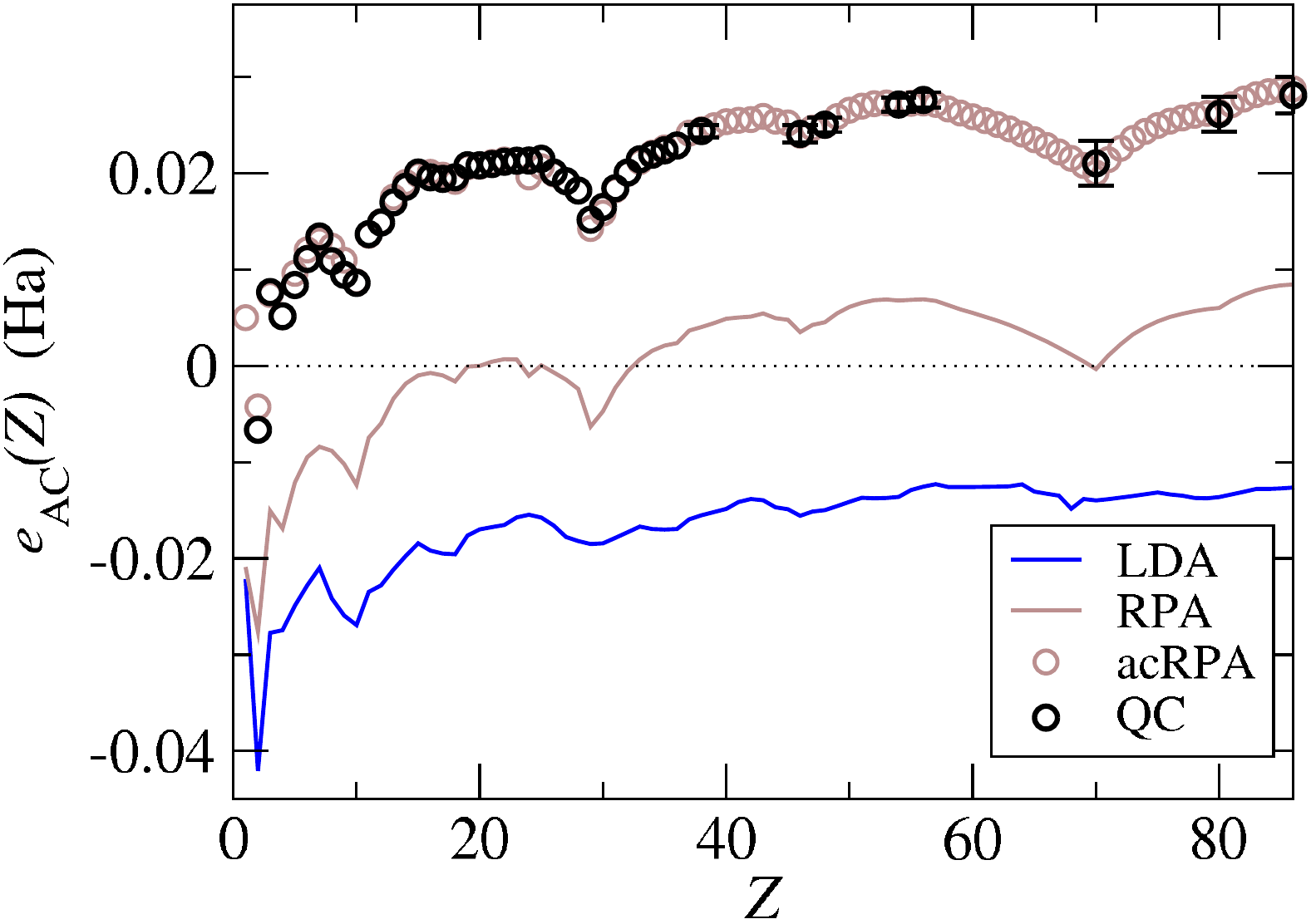}
\caption{Asymptotically corrected correlation energy per electron 
[Eq.~(\ref{eac})]
versus $Z$ for 
accurate quantum chemistry (QC) calculations (black circles), 
the RPA (brown solid line), corrected RPA data (brown circles),
and LDA.
\label{apEcall}
}
\end{figure}
The locality conjecture, applied to correlation alone, implies that
the  leading term of Eq.~(\ref{aclda}) is exact.  In fact, this has been proven for atoms relatively
recently,\cite{KR10} which shows definitively that the curves in Fig. \ref{Ecall}
do not saturate.  It also suggests we define:
\ben
e\ac(Z) = E\c/Z + A\c \ln Z
\label{eac}
\een
as the asymptotically-corrected correlation energy per electron. This is plotted
in Fig. \ref{apEcall}.  Now the curves do appear to saturate, although
the oscillations across open shells make accurate extrapolation of the
large-$Z$ limit very difficult.
For any such curve, we define
\ben
B\c = \lim_{Z\to\infty} e\ac(Z),
\een
and try to estimate this value as accurately as possible.  
This will be the subject of the next section.

\begin{figure}
\includegraphics[width=0.95\linewidth]{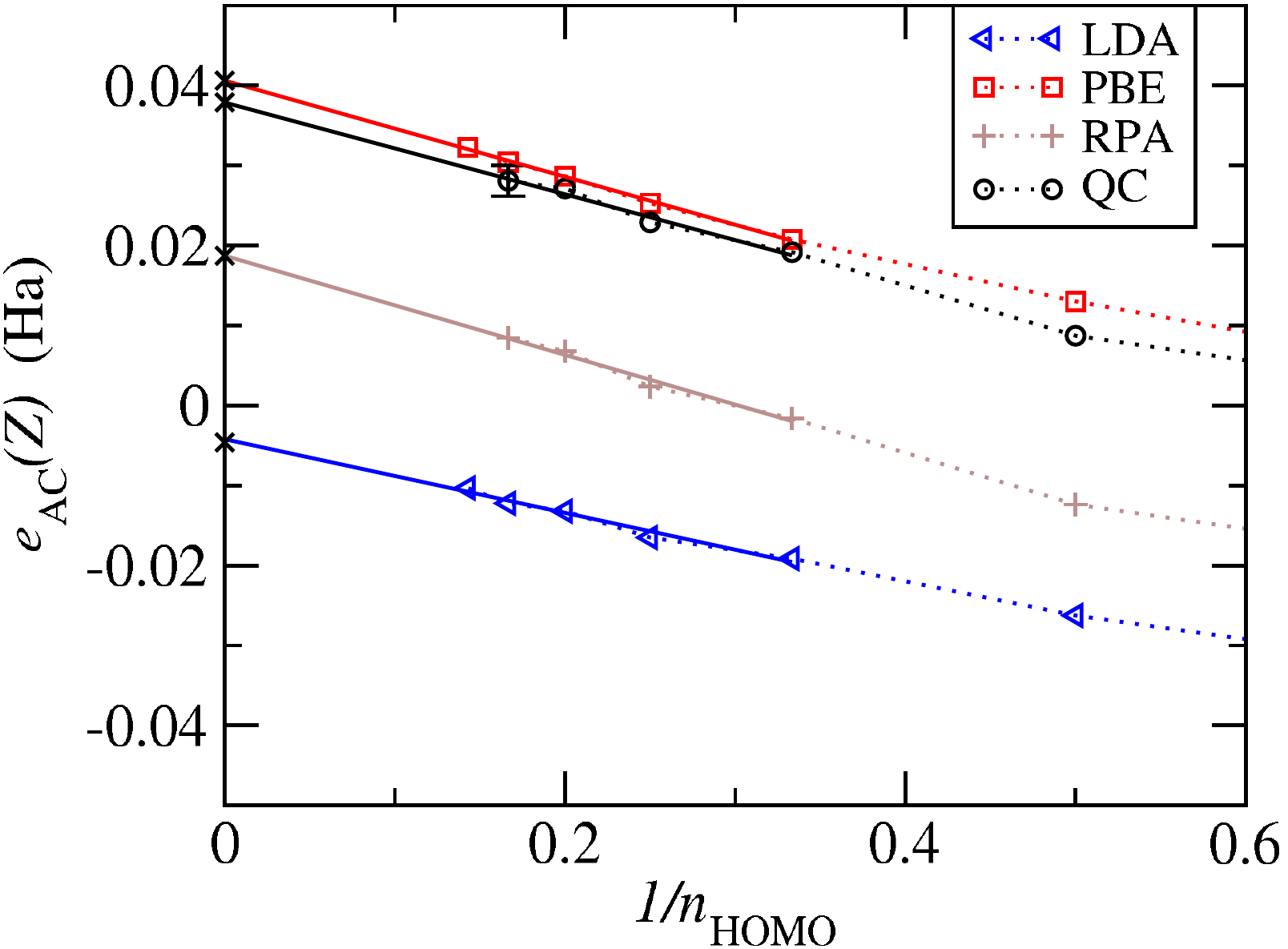}
\caption{\label{napEcall}
Asymptotically-corrected correlation energy per electron as a function
of inverse highest occupied shell for noble gases, for accurate
quantum chemistry data (QC) and within RPA, LDA and PBE.
For QC, the error for Rn is shown, others are smaller than 
data point size.
}
\end{figure}
\begin{figure}
\includegraphics[width=0.95\linewidth]{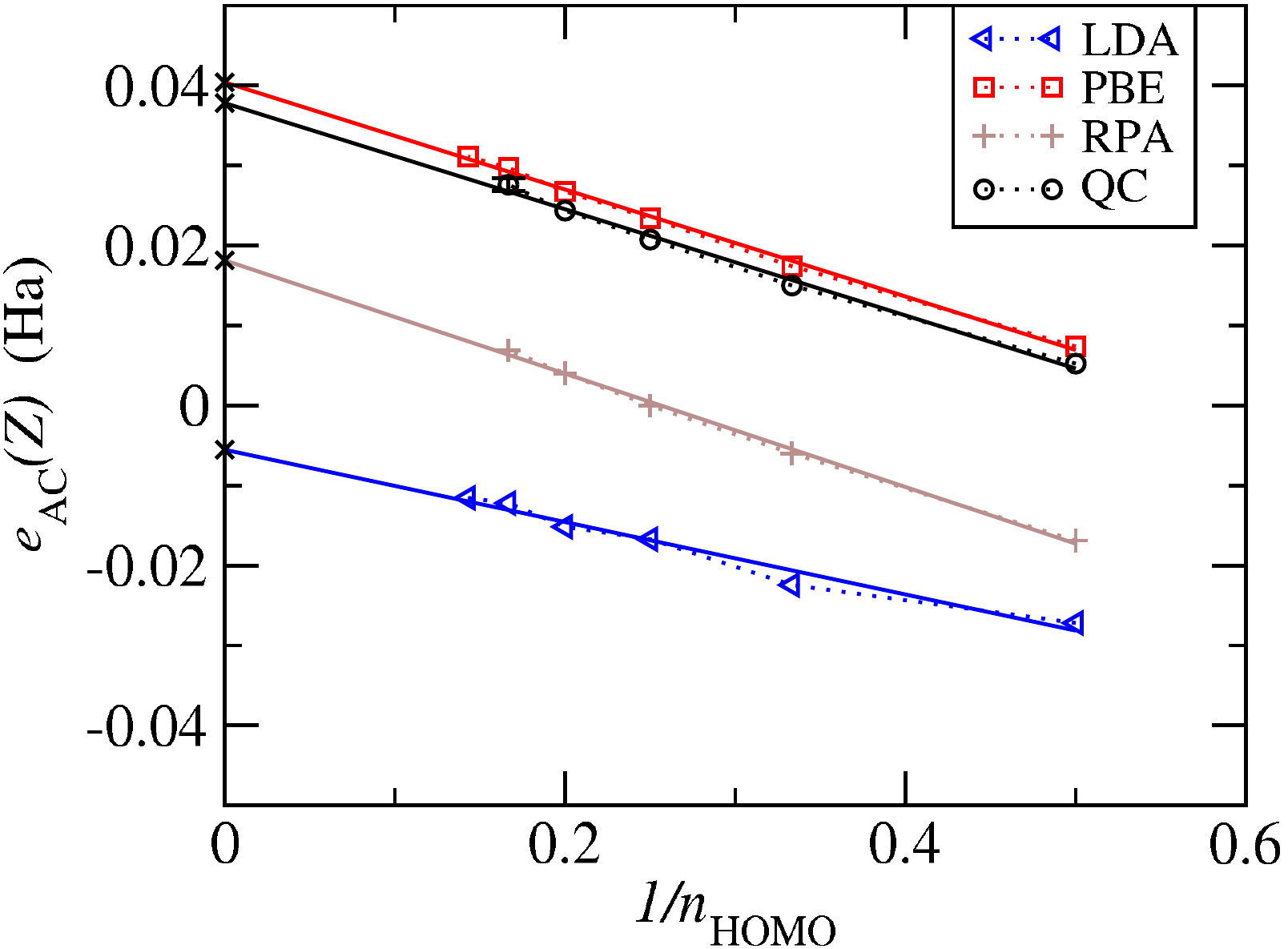}
\caption{\label{AEapEcall}
Same as Fig. \ref{napEcall}, but for alkali earths.
}
\end{figure}
But to get a quick idea, using techniques that have worked for exchange,\cite{EB09,LCPB09}
we largely eliminate the effect of shell structure by taking
only noble gas atoms, and plot as a function of $1/\nh$, the principal
quantum number of the highest occupied shell, as in Fig. \ref{napEcall}.  
Clearly, the RPA and LDA values of $B\c$ are quite different from the
reference QC value.
(The PBE correlation functional\cite{PBE96} is included, as its asymptotic
behavior is almost perfectly parallel to the benchmark data, and  we use this
in the next section to improve our extrapolation.)
Repeating the procedure for alkali earths shows that almost identical
results occur, as shown in Fig. \ref{AEapEcall}.
Fitting noble gas data to the form
\ben
e\ac\nob(\nh) = B\c\nob - C\c\nob/\nh,
\label{linearnhomofit}
\een 
we find
$B\c\nob \!=\!  0.0371$ and $C\c\nob\!=\! -0.0536$.  Repeating the fit for
the alkali earths 
we find 
$B\c\AE\!=\! 0.0378$ and $C\c\AE\!=\! -0.0644$, reflecting a consistency
in the value of $B\c$ extracted from either series.

\def\nsm{\tilde\n\nob}
Finally we can convert our numerical extrapolation versus $\nh$ 
into a smooth function $e\ac\nob(Z)$
in order to explore trends across the entire periodic table.
To do so, first notice that 
the data oscillate slightly around each fit, with every other point
being above or below.  This reflects the double-step in the periodic
table between the appearance of new rows.
To account for this,  while still analyzing 
atoms down a given column, we define
\ben
\nsm(Z)= (6 Z + 8)^{1/3}-2,
\label{nsm}
\een
which smoothly interpolates $\nh$ between even and odd noble gas atoms,
and is asymptotically correct as $Z\to\infty$.\cite{footnote2}
We can now write
\ben
e\ac\nob(Z) = B\c\nob - C\c\nob/\nsm(Z),
\label{linearfit}
\een 
an estimate of $e\ac$ for all $Z$.
To repeat the process for the alkali earths, 
we use $\nsm(Z-2)+1$ instead of $\nsm(Z)$ in Eq.~(\ref{linearfit})), 
and the AE coefficients $B\c\AE$ and $C\c\AE$.
The two asymptotic energies converge rapidly onto each other for large-$Z$,
as shown in Fig.~\ref{fig:apEcallfit}.

%

Having identified that $B\c$ appears to approach a finite value, in the
section following we make a variety of constructions in order to estimate its value
as accurately and reliably as possible from the available data.
We also use some of the same techniques to fill in large-$Z$
non-spherical atomic correlation
energies in Tables~\ref{tab:EcI} and \ref{tab:EcII}.

\ssec{Testing extrapolation methods on LDA}

As an obvious test of our numerical extrapolation methods, we consider self-consistent LDA
calculations for neutral atoms. 
Because such calculations are of irrelevant computational cost, 
we 
enlarge the data set up to $Z=460$.  This is the 12th row of the non-relativistic periodic
table, assuming Madelung's rule continues indefinitely.  Note that any calculations beyond
$Z=100$ are simply a mathematical device to reduce the range of $Z$ over which we extrapolate.

\def\HDLDA{^{\rm HDLDA}}
\begin{figure}
\includegraphics[width=0.95\linewidth]{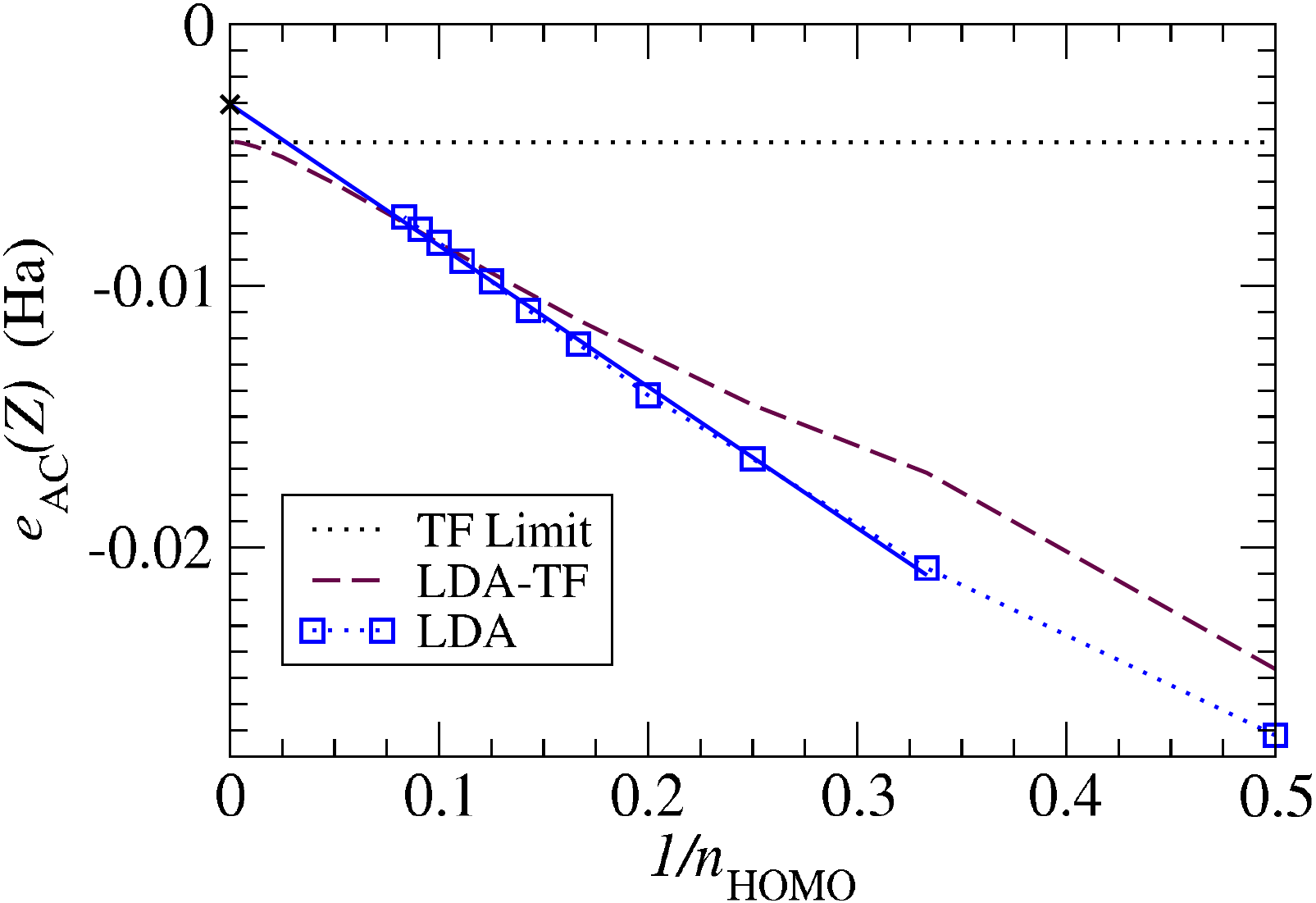}
\caption{\label{Ecasylda}
Asymptotically-corrected correlation energy per electron versus $1/\nh$. 
Squares show self-consistent LDA energies for the alkali earth 
Be ($\nh\!=\!2$) and also the average of noble gas and alkali earth for 
$\nh\!\ge\! 3$.  
Solid line is linear fit extrapolating 
$\nh \!\to\! \infty$, and dashed shows LDA evaluated with 
asymptotic Thomas-Fermi density.
The horizontal line is at the analytic value of $B\c\LDA$.
}
\end{figure}

As we go to larger $Z$, we find even/odd oscillations becoming less 
problematic; they can be largely eliminated by 
averaging over the noble gas and the alkali earth atom in each row.
These are both 
spherical (and so appear in our original QC data set) and not spin polarized.  
In Fig. \ref{Ecasylda}, we plot the averaged results of self-consistent LDA
calculations (squares connected by dotted line), 
the linear extrapolation of these to the asymptotic limit (solid line), 
and mark the analytical form of this limit from Eq. (\ref{eqBLDA}).
The LDA energy evaluated on the TF density for each value of $Z$ 
(dashed line) is also shown.

One can see that the na\"ive linear trend of $e\ac$ with $1/\nh$
must eventually fail for $1/\nh < 0.1$.  This occurs even when the 
TF density is used, due to the logarithmic terms in the high-density
expansion of $\epsilon\c\unif$.
The difference between the self-consistent and TF curves also 
appears to be non-linear in this region.  To understand this, 
we expand the uniform gas energy in the high-density limit to the
next two orders beyond Eq. (\ref{epsunif}):
\ben
\epsilon\c\LDA = c_0\ln r_s - c_1 + c_2\, r_s\ln r_s - c_3\, r_s, 
r_s \to 0,
\label{HDLDA}
\een
where 
$c_2 \!=\! 0.0066$ and $c_3 \!=\! 0.0104$ are known from perturbation theory.\cite{PW92}
Inserting the TF density yields
\ben
e\ac\LDA[\n\TF](Z)\to 
B\c\LDA\, +(\tilde C\c\, \ln Z + \tilde D\c)/Z^{2/3}
\een
where
\ben
\tilde C\c = -\frac{2}{3} c_2 \langle r_s \rangle = -0.051,
\een
and
\ben
\tilde D\c =\left[c_3 \langle r_s \rangle + c_2
                      \langle r_s \ln(r_s) \rangle\right] = 0.207.
\een
In these equations, $r_s \!=\!  a (3/f(x))^{1/3}$ is the local Wigner-Seitz radius
of the TF density for $Z=1$, and 
\ben
\langle g \rangle = \int_{-\infty}^\infty dx\, x^2 f(x)\, g(x).
\label{TFmoment}
\een
The important feature is that this expansion is ill-behaved.  
The culprit is the nature of the average over 
TF density [Eq.~(\ref{TFmoment})], as $x^2f(x)[r_s(x)]^n$ decays only 
as $x^{2n-4}$ as $x\to\infty$.
As a consequence, these next two terms, of order $n\!=\!1$,
have much larger coefficients than the leading, order $n\!=\!0$
terms of Eq. (\ref{epsunif}), 
and the following two terms 
diverge on the TF density.  
This poor behavior is inherited from
the uniform gas and shows how unsuitable all but the leading terms of the LDA are for
approximating the correlation energy of atoms.  Furthermore, it causes the
structure we see in the figure as $Z\to\infty$.
Thus, the LDA curve is especially difficult to extrapolate accurately.
Nonetheless, performing 
a naive application of the linear extrapolation for $\nh \geq 3$ yields
-3.0~mHa, an underestimate of 1.5 mHa from the analytic value.  
Thus we expect no larger error from our extrapolation process.

\ssec{Accurate estimate of $B\c$}

Finally, we perform the extrapolation that produces our best estimate for $B\c$.
To achieve the maximum accuracy, we compare the accurate QC data with PBE results.\cite{PBE96}
The primary reason is that the PBE functional was constructed to yield an accurate finite result
in this limit, by correcting the LDA energy.  Our use here is simply that we can extrapolate the
difference between QC and PBE, while extracting the PBE result analytically.  
The parallel trends in $\nh$ for 
the PBE and QC in Figs.~\ref{napEcall} and~\ref{AEapEcall} imply
that their difference tends to a constant.
Thus the resulting fit ought to be almost exclusively
a measure of $B\c$, with minimum influence from the details of the fit.

\def\blPBE{^{\rm BL,PBE}}
The PBE beyond-local correlation energy for spin-unpolarized densities
is\cite{PBE96}
\ben
E\c\blPBE[\n] =  \int d^3r\, \n(\br)\ h\c\PBE(r\s(\br),t(\br)),
\label{EcPBE}
\een
where
$t\!=\!  |\nabla \n|/(2 k_s \n)$ is the dimensionless gradient
for correlation, 
$k\s\!=\! 2 (3\n/\pi)^{1/6}$ is the TF screening
length, and
\ben
h\c\PBE(r\s,t)= c_0\, \ln \left(
1 +\frac{\beta}{c_0}\, t^2\, f\c(\tilde A t^2)\right),
\label{Hpbe}
\een
where
\ben
f\c(y)=(1+y)/(1+y+y^2), 
\een
\ben
\beta \tilde A^{-1} = c_0\left[ \exp (-\epsilon\c\unif/c_0) -1\right],
\label{Adef}
\een
and $\beta\!=\! 0.066725$.
This peculiar combination of functions results from requiring it to reduce to
LDA for uniform densities, yield a finite value in the high-density limit of
finite systems, and, when combined with PBE exchange, recover the LDA linear response
for a uniform electron gas.\cite{PBE96}  It is the first two conditions
which ensure that it yields a finite value for $B\c$ that differs from LDA.

$B\c\PBE$ is found by considering the contributions linear in $Z$ generated by
replacing Eq.~(\ref{nTF}) in the high-density large-$Z$ limit, where
$\tilde A \to 1/r\s$, $y\to 0$, and $f\c\to 1$.  
As a result,
\ben
\Delta B\c\PBE = c_0\, I_3(\beta/c_0),
\een
where
\ben
I_3(u)=\int_0^{\infty} dx\, x^2\, f(x)\, \ln \left( 1+ \frac{u}{\sqrt{6}} \frac{df/dx}{f^{7/6}(x)} \right),
\een
is a well-defined integral over the TF density.\cite{LCPB09}
For $u\!=\! \beta/c_0\!=\! 2.146119$, we find $I_3\!=\! 1.411164$, yielding
\ben\label{eqBPBE}
B\c\PBE = 0.03936.
\een
in agreement with Ref.~\onlinecite{PCSB06}.

\begin{figure}
\includegraphics[width=0.95\linewidth]{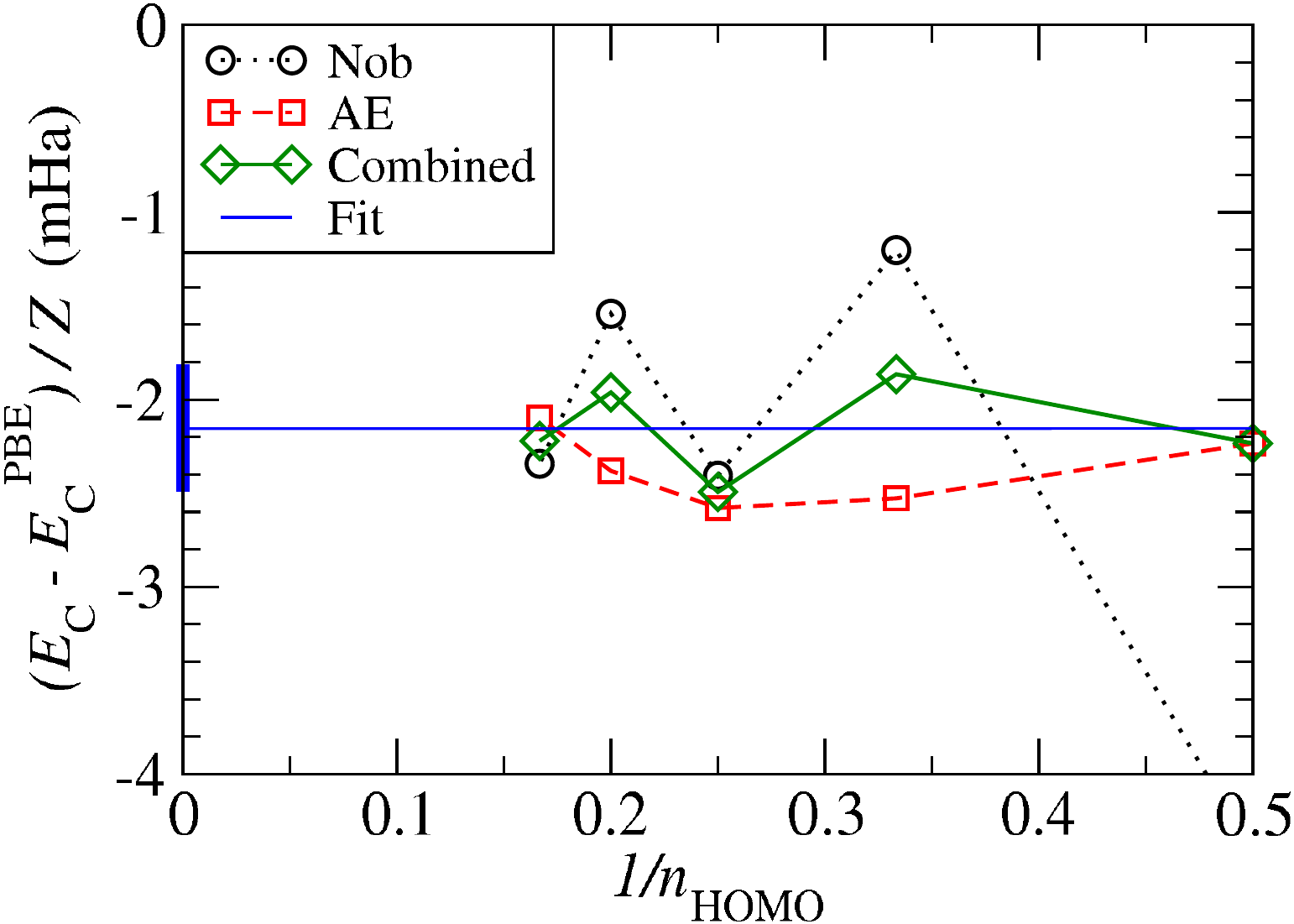}
\caption{\label{colaverages}
Difference between the correlation energy-per-electron of QC data and
the PBE for noble gas (Nob) and alkali earth (AE) atoms, 
plotted versus $1/\nh$, along with
the combined averages for $\nh\!=\! 3$ through $\nh\!=\! 6$.  
Linear fit [Eq.~(\ref{nicefit})] in blue,
with error bar in the $1/\nh \!=\! 0$ asymptote shown as solid thick bar.
Units in mHa. 
}
\end{figure}
In Fig. \ref{colaverages}, we plot the energy difference in mHa between QC and PBE
per electron.  Because the difference is so small, on this scale, the differences between noble
gases and alkali earths are very visible, especially for odd rows.  But their mean is seen to
converge nicely.  A fit yields
\ben
(E\c - E\c\PBE)/Z = -0.00215(34) + 0.0000(11)/\nh.
\label{nicefit}
\een
Combining the analytic result of Eq. (\ref{eqBPBE}) with this one,
we conclude that the
asymptotic value of $B\c$ is 37.2~mHa.
\begin{table}[h]
\begin{tabular}{|c||c|c|c|}
\hline
    &  AE~~~   &    noble~~~~  & analytic~~ \\
\hline \hline
LDA & -5.9(16) & -5.0(1.1)  & -4.51 \\ 
PBE & 40.2(7)   & 40.2(5)   &  39.3 \\     
RPA & 18.1(7)   & 18.7(1.4) &  15.2 \\ 
QC  & 37.8(9)   & 37.1(1.6) & N/A  \\
\hline
\end{tabular}
\caption{
Values for $B\c$ in different approximations via numerical extrapolation
and analytically.  Reported errors, in parentheses, are 
statistical errors in the linear regression used to make the extrapolation.
Errors in extrapolation can also be judged from the DFT
cases where analytic results are known.  The QC values are consistent with
the best estimate of 37.2 mHa from Fig.~\ref{colaverages}.
}
\label{tab:Bc}
\end{table}
Table \ref{tab:Bc} compares the different estimates, showing their
consistency, and suggesting the extrapolation
error is no more than 1.5mHa.

\ssec{Corrections to RPA}

There is considerable current interest in using RPA and beyond-RPA methods for quantum
chemistry.\cite{BF13}  While many good features are known,
a major drawback to pure RPA calculations for thermochemistry is the inaccuracy
of RPA atomic energies, and its impact on atomization energies.  

\begin{figure}
\includegraphics[width=0.95\linewidth]{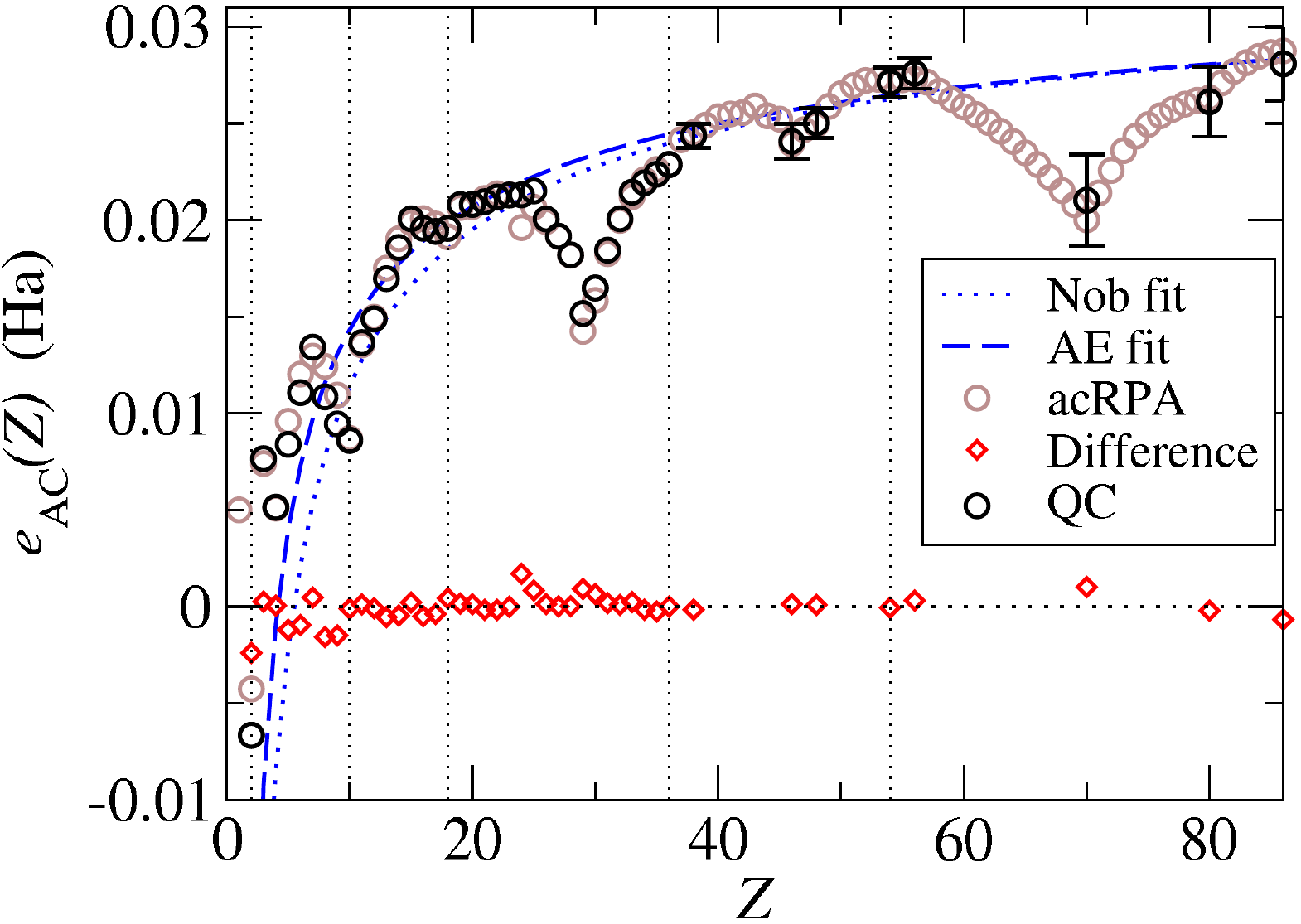}
\caption{\label{fig:apEcallfit}
Asymptotic correction of RPA data: black points are QC data, brown are RPA data corrected via Eq. (\ref{dEcRPA}), and red, their difference.
Also shown as dashed and dotted lines, smooth asymptotic curves
$e\ac\AE(Z)$ and $e\ac\nob(Z)$
defined through Eq.~\ref{linearfit} and text following.
}
\end{figure}
Unfortunately, the quantum chemical data becomes sparse for large $Z$, being
limited to spherical atoms.
However, our RPA calculations are less limited, and their error relative to 
the quantum chemical data set is relatively simple.
We fit the differences between the RPA correlation energies and
the QC ones for noble-gas atoms using:
\ben
\Delta E\c\RPA/Z = 0.0199 + 0.00246/\nsm(Z),
\label{dEcRPA}
\een 
where $\nsm(Z)$ is 
given by Eq.~(\ref{nsm}).
We denote by acRPA the RPA energies with the correction of 
Eq. (\ref{dEcRPA}) added to them.
The results are presented in Tables~\ref{tab:EcI} and \ref{tab:EcII}.

In Fig. \ref{fig:apEcallfit}, we plot both the accurate correlation energies 
and the acRPA values, and the residual error.
Clearly, the fit is extremely accurate and errors become negligible for large 
enough $Z$.  Finally, we note that the constant contribution to the fit, 
19.9 mHa, is close (but different from) the error RPA makes for the correlation 
energy per electron in the high density limit of the uniform gas, 24.2 mHa, 
as discussed in Section \ref{theory}.  
We can understand one contribution to this difference by considering the
PBE for RPA correlation energies developed by Kurth and Perdew.\cite{Kurth1999}
The expression of the corresponding enhancement factors
is reported in Eq. (27) of Ref.~\onlinecite{Zidan2000}, from which we readily obtain
\ben\label{eqBRPA}
B\c\PBERPA = 0.01518,
\een
the difference with PBE being again due to RPA correlation energy of the uniform gas.
Since the PBE construction uses the value of  the gradient
expansion coefficient $\beta$ from the high-density limit in the RPA approximation,  
the {\em gradient} contributions to $B\c\PBE$ and  $B\c\RPA$ are the same.
Assuming this procedure is highly accurate for approximating RPA, the remaining
difference is only 3 mHa - well within the numerical error bars
for RPA correlation energies, estimated to be up to 4.6~mHa (3\% of 84~mHa) for
$Z\!=\! 86$.\cite{footnote3}

Note that this correction formula concerns the relation between an 
approximation to the many-electron problem and the exact result, and is 
entirely independent of any density functional analysis. Its value is that we 
can create an enhanced data-set of the QC data, plus RPA-corrected
estimates for the larger open-shell atoms. 
We use acRPA for benchmark values wherever QC values are unavailable, i.e., 
the non-spherical atoms of Table \ref{tab:EcII}.
Our simple atomic correction, Eq. (\ref{dEcRPA}),
is remarkably accurate, as shown in Tables
\ref{tab:EcI} and \ref{tab:EcII}, with few errors greater than 1 mHa.
A scheme which reproduces this correction, but also works well for molecules at equilibrium
bond lengths, would overcome the limitations of pure RPA.  We suggest testing existing schemes for their
behavior in the large-$Z$ limit, characterizing them by the value of $B\c$.

\ssec{Deviations from smooth functions of $Z$\label{IIIF}}

\begin{figure}[htbp]
\includegraphics[width=\columnwidth]{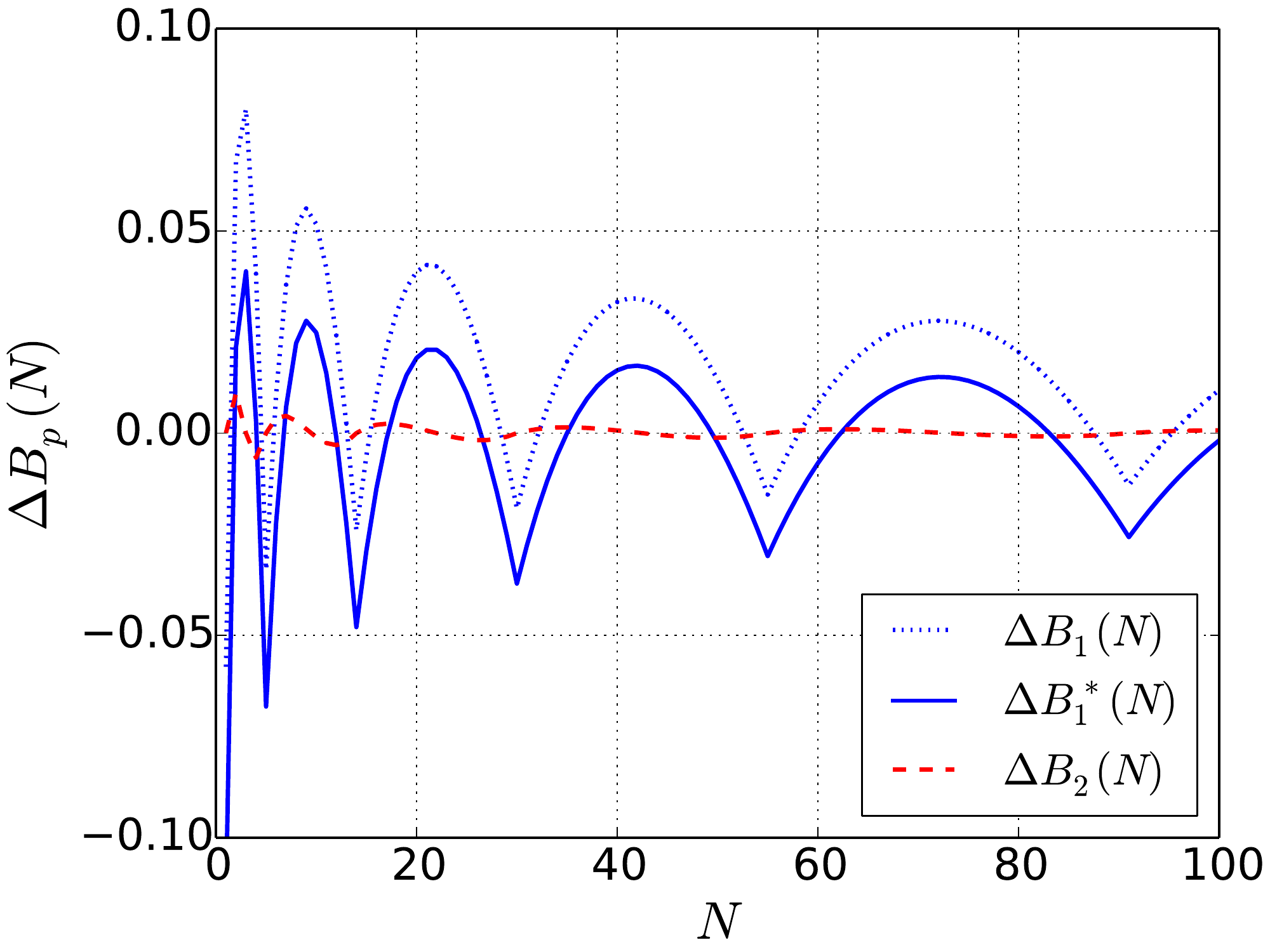}
\caption{Errors $\Delta B_p(N)=B_p(N)-S_N$ ($N\geq 1$)
in partial sums $B_p(N)$ [Eq.~(\ref{Bp})] to different orders 
for the mathematical series from the introduction.
The blue solid curve ($B_1^*$) has the mean of $b_2(N)$ subtracted,
to emphasize the shell structure.}
\label{dB}
\end{figure}
To motivate this final section, we return
to the didactic example from the Introduction.
In Fig.~\ref{dB} we show (blue dotted line) the error in the asymptotic 
curve $B_1(N)$ derived in that section and plotted in Fig.~\ref{sA}.
Subtracting the mean value of $b_2(N)$ across a shell,
$[-12 (3N)^{1/3}]^{-1}$, yields the error on $B_1^*$ shown as a
blue solid line. 
This error is \text{not} a smooth function of $N$
but has cusps at shell boundaries, 
is periodic in $\nu$, the fraction of the shell that is filled, and only 
slowly decaying with shell number.  
Careful analysis~\cite{E88} yields
\ben
b_2(N)=\frac{1-12\nu(1-\nu)}{12 (3N)^{1/3}},
\label{b2}
\een
and addition of this term to form $B_2(N)$ yields the error
shown as red dashes.
The errors between shells are reduced by an order of magnitude relative
to $B_1(N)$, and
the error for $N\!=\! 1$ is a mere $3\times 10^{-5}$, i.e., 30 $\mu$H.
Thus it behooves us to
look    
at the structure of such terms.

Of course, exchange and correlation of real atoms is much more complicated,
and we do not have access to the exact asymptotic expansion for these
quantities.  
But as our example shows, we may 
numerically illustrate
the nature of the underlying higher-order asymptotic corrections simply by 
showing the deviation of our data from smooth behavior. 
To do so, we first
obtain smooth curves from our analysis of QC data
from single columns of the periodic table, equivalent to the 
leading order terms of our didactic example.
Secondly, we plot the difference between smooth asymptotic curve and
QC data augmented by acRPA estimates to obtain a sense of the higher-order 
terms that determine shell structure along rows of the periodic table.
We have already constructed such curves for correlation:
$e\ac\nob(Z)$ fitting values for the noble gases using
Eq.~(\ref{linearfit}) and an analogous curve, $e\ac\AE(Z)$, for the alkali 
earths.  
As shown in Fig.~\ref{fig:apEcallfit},
the two curves have the same behavior at large $Z$, but 
the AE curve is more reliable for lower $Z$, so we adopt the latter
as standard.

To create smooth
functions for exchange, 
we fit exchange energy results with 
continuous functions of $Z$, and apply these functions for every value of $Z$.
We standardize on alkali earths to be consistent with correlation.
As we know LDA becomes relatively exact for exchange for large $Z$, we define
\ben
e\ax=E\x/Z-A\x\, Z^{2/3},
\een
and fit EXX energies for alkali earths with 
\ben
e\ax\smooth= -(0.1466 x^3 +0.0314 x^2 +0.0823)
\label{eaxsmooth}
\een
where $x\!=\! 1-Z^{-1/3}$.  
This gives $B\x \!=\!  -0.260$ in agreement with a previous
estimate of -0.24.\cite{EB09}  The value of -0.08 at $Z\!=\! 1$ is
somewhat off the actual value of -0.11, but gives a close fit to $e\ax$
for all  $Z>2$.

\begin{figure}[h]
\includegraphics[width=0.9\linewidth]{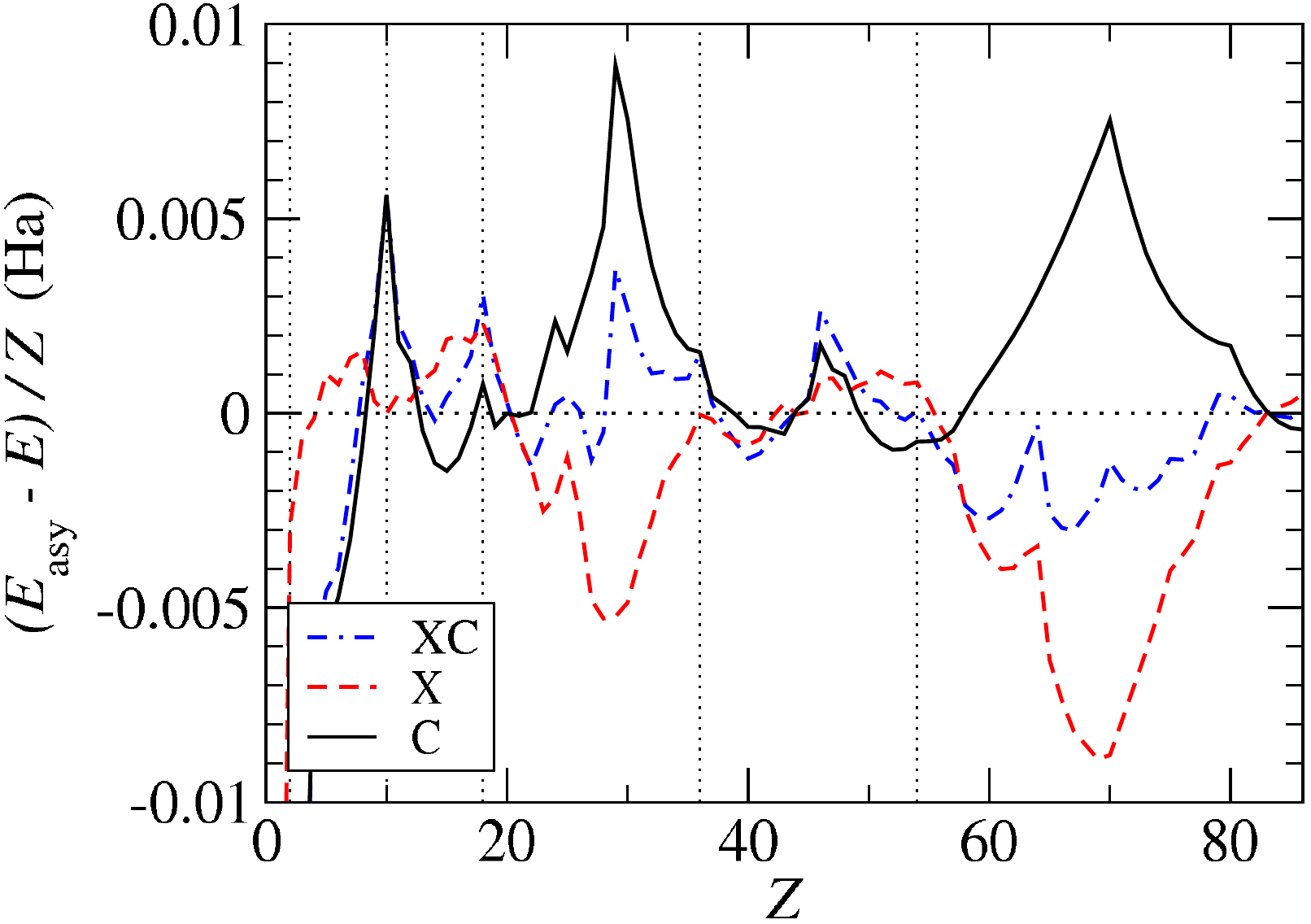}

\caption{\label{Tents}
Difference between the smooth asymptotic limiting functions for 
asymptotically corrected exchange and correlation energies and their 
QC values, versus atomic number.  XC is exchange-correlation.
Vertical lines denote location of noble
gases to better visualize the location of periodic table rows.
}
\end{figure}
In Fig \ref{Tents}, we plot the deviations of the exchange and correlation energies
per electron from the smooth fits.  This is where acRPA is important:  To
see the periodic pattern in the deviations in the correlation energy by
including large $Z$ values.  We have also included vertical lines at the
noble gas atoms, to show the periodicity.

There are many interesting features in this plot.
\bei

\item There is a clear (anti-)correlation between the correlation deviation
and the exchange deviation, showing tremendous cancellation of `errors'
in the separate components of the deviations.  Thus, adding
Eqs (\ref{eaxsmooth}) and $e\ac\AE(Z)$ yields a much more accurate
estimate of XC than either separately.  Shell structure is much greater
in either exchange or correlation alone than in XC.

\item The greatest deviations occur about the middle of the odd rows,
and deviations are almost negligible throughout the 4th row.

\item Filled angular momentum subshells tend to be marked by
cusp-like behavior, especially for correlation.  These are more
apparent if the irregularities in filling for d and f subshells are ignored.

\item We observe the pattern that whenever the most loosely
bound electron also has the highest angular momentum, the deviation
grows as that angular momentum subshell is filled, and drops once the next subshell is
filled.  For example, for $Z\!=\! 21$ to $29$, the 3d-shell is being filled
and the deviation grows, and then drops rapidly as the 4p's are filled.  
Note that this only happens when a given subshell is filled for the first
time; there is little effect when the 4d's are being filled.  A similar
pattern appears for the 4f's, and (to a lesser extent) even for the 2p's.

\eei

\sec{Consequences}

The majority of the numerical work in this manuscript has been devoted to the
extraction of an accurate numerical estimate for $B\c$, which characterizes
the leading correction to LDA correlation for atoms, and is strong evidence for
the locality conjecture.  In this section, we assume that conjecture is true
and discuss its consequences for DFT approximations.  We use the mathematical
prolog and its relevance to the kinetic energy of non-interacting atoms to 
illustrate this, by discussing $T\s[\n]$, $E\x[\n]$, and $E\c[\n]$.

\ssec{Relevance of uniform gas to Coulombic matter}

In the math prolog, we showed that, for the kinetic energy of non-interacting atoms:
\ben
T\s(Z) = (3/2)^{1/3}\, Z^{7/3} - Z^2/2 + \dots
\een
For real atoms,
\ben
T\s(Z) = A\s\, Z^{7/3} - Z^2/2 + \dots,
\een
where $A\s\!=\! 0.768745$ is given exactly by TF theory,
while
\ben
E\x(Z) =  -A\x\, Z^{5/3} + B\x Z + \dots,
\een
and
\ben
E\c(Z) = -A\c\, Z \ln Z + B\c Z + \dots
\een
In every case, a local density approximation yields the leading term for large $Z$.  
Thus, LDA yields the exact large-$Z$ limit of each component for {\em every} electronic
system.  It is a {\em universal} limit for all matter.  
The form and coefficients of local approximations can be most easily calculated from
the uniform electron gas, but in principle can be extracted from taking any system, such
as atoms, to the large-$Z$ limit.

But note that correlation is different from exchange and the kinetic energy.  In the latter two,
the corrections to the dominant term are smaller by a factor of at least $Z^{-1/3}$.  Thus the corrections
are relatively small for $Z > 50$, and relatively easily identified.  For correlation, the
leading term is only $\ln Z$ larger, meaning $Z$ must be vast before it dominates.  Thus, lists of
values of $E\c(Z)$ for $Z < 100$  {\em alone}  cannot  be used to extract this
behavior accurately.  
Moreover, approximations (especially any
designed only for lighter atoms) can ignore this contribution and still remain accurate.  Moreover,
since that term is determined by the high-density behavior of the uniform gas, and the next term
in LDA is not accurate, all the rest of the correlation energy of the uniform
gas is not particularly relevant to atomic correlation energies.  Thus almost all popular approximations
to $E\x$ reduce to LDA for uniform densities, but not so for $E\c$.  

Except for model 1-d systems,\cite{ELCB08,CLEB10,CLEB11,ECPG15,RLCE15} 
no-one has ever written down a general functional
approximation that recovers the leading corrections to LDA for all systems.  The 
gradient expansion approximation does this for slowly-varying densities, but then is quite
incorrect for atoms.\cite{EB09}  Generalized gradient approximations attempt to capture both, but find
they are irreconcilable.\cite{ZY98,PBE98}

\ssec{Performance of semilocal approximations}

The locality principle claims that one can understand the successes and failures of semilocal
DFT in terms of the large-$Z$ expansion.  We now consider three basic properties often 
calculated with such approximations:  ionization potentials, bond energies, and bond lengths.
Note that all these depend only on energy differences, not the total energies studied here.
But we argue below that the qualitative behavior of semilocal approximations for each of
these cases can be understood in terms of the locality principle.

The simplest case is ionization energies, which can be studied for atoms.\cite{CSPB10}
Within numerical
accuracy, at the exchange level, LDA yields the exact ionization energy in the large-$Z$ limit,
and GGA (in the PBE case, at least) correctly reduces to LDA in this limit.  
This is consistent
with the fact that GGA's and global hybrids
do not typically yield significant improvement
over LDA for ionization energies.\cite{PCVJ92,CPR10}  An analogous result was found for XC: PBE corrections
to LDA ionization energies
are very small (or zero) in this limit.  Thus, at least for atoms, LDA becomes relatively
exact for ionization potentials in this limit, and our standard GGA's do not yield much 
improvement.  They do not produce the leading correction in the asymptotic expansion for the
ionization potential, and so are not systematically better than LDA for real systems.

Next consider bond energies.  GGA's typically improve atomization energies relative
to LDA.   Both PBE and B88 yield accurate asymptotic corrections to LDA exchange for large $Z$.
Thus the corrections are accurate for both the equilibrium molecule and the isolated atoms,
sufficiently so as to produce improved energy differences.   This suggests that both cases
are improved by improving the asymptotics.

But now consider what happens as a molecule is stretched relative to equilibrium. 
The molecular levels become nearly degenerate.  The smaller the gap, the further they are away from
the large-$Z$ limit, which assumes a continuum of levels.  To see this semiclassically, we note
that the Coulson-Fisher point in LDA for H$_2$ is at $R\!=\! 3.3$~a.u.\cite{BA96}  If we consider the
point where the HOMO level of the exact KS potential just touches the maximum 
in the KS
potential well, this occurs at about the same distance.\cite{BBS89}  
As $R$ increases through 3.3, the classical
turning point surface of the KS potential at the HOMO energy divides into two isolated
regions.  Thus the onset of strong static correlation coincides with a qualitative change
in the nature of the semiclassical (i.e. local) approximation.  The asymptotic expansion works
well at equilibrium or for isolated atoms, but fails completely as 3.3 is approached,
because of the degeneracy.  This is the semiclassical explanation of the failure of semilocal
approximations as bonds are stretched.

Finally, we discuss bond lengths, which are determined by the derivative of the binding energy
curve at equilibrium.  The asymptotic expansion in $Z$ should work (almost) equally well for
separations in and near equilibrium.  The classical turning point structure at the HOMO level
does not change qualitatively.  Furthermore, the asymptotic behavior for large-$Z$ remains the
same for each value, and so irrelevant for the derivative.  Exchange GGAs that capture
the correct asymptotics of the total energy violate the gradient expansion, which should be
accurate for these small changes.  This was (part of) the reasoning behind PBEsol, which restores
the gradient expansion for exchange and improves lattice constants of many solids.\cite{PRCV08}

\sec{Conclusions}

The central result of this paper is to combine existing numerical evidence for 
the correlation energy of non-relativistic atoms,
together with knowledge of the asymptotic form for the dependence of 
this energy on $Z$,
to extract the leading 
correction to the local density approximation for atoms in the large-$Z$ limit.
This is a key number that characterizes
the leading error made by LDA, and that can be corrected by more sophisticated approximations.
We have argued that this number is non-empirical, and should be used as an exact condition
for non-empirical construction of functionals.  

In a preliminary report, we constructed a
non-empirical GGA that was asymptotically correct for atoms.\cite{BCGP14}   The method was then
used to impose exact conditions on the SCAN meta-GGA.\cite{SRP15}

\usec{acknowledgments}
This work was supported by NSF CHE-1464795.  
SP was supported by the European Community through the FP7's MC-IIF MODENADYNA, grant agreement No. 623413 and
by the European Commission (Grant No. FP7-NMP-CRONOS). 
TG recognizes computing support from the Griffith University Gowonda
HPC Cluster.
We thank John Perdew and Raphael Ribeiro for many useful discussions and
Eberhard Engel for use of his atomic DFT code, OPMKS.


\begin{thebibliography}{92}%
\makeatletter
\providecommand \@ifxundefined [1]{%
 \@ifx{#1\undefined}
}%
\providecommand \@ifnum [1]{%
 \ifnum #1\expandafter \@firstoftwo
 \else \expandafter \@secondoftwo
 \fi
}%
\providecommand \@ifx [1]{%
 \ifx #1\expandafter \@firstoftwo
 \else \expandafter \@secondoftwo
 \fi
}%
\providecommand \natexlab [1]{#1}%
\providecommand \enquote  [1]{``#1''}%
\providecommand \bibnamefont  [1]{#1}%
\providecommand \bibfnamefont [1]{#1}%
\providecommand \citenamefont [1]{#1}%
\providecommand \href@noop [0]{\@secondoftwo}%
\providecommand \href [0]{\begingroup \@sanitize@url \@href}%
\providecommand \@href[1]{\@@startlink{#1}\@@href}%
\providecommand \@@href[1]{\endgroup#1\@@endlink}%
\providecommand \@sanitize@url [0]{\catcode `\\12\catcode `\$12\catcode
  `\&12\catcode `\#12\catcode `\^12\catcode `\_12\catcode `\%12\relax}%
\providecommand \@@startlink[1]{}%
\providecommand \@@endlink[0]{}%
\providecommand \url  [0]{\begingroup\@sanitize@url \@url }%
\providecommand \@url [1]{\endgroup\@href {#1}{\urlprefix }}%
\providecommand \urlprefix  [0]{URL }%
\providecommand \Eprint [0]{\href }%
\providecommand \doibase [0]{http://dx.doi.org/}%
\providecommand \selectlanguage [0]{\@gobble}%
\providecommand \bibinfo  [0]{\@secondoftwo}%
\providecommand \bibfield  [0]{\@secondoftwo}%
\providecommand \translation [1]{[#1]}%
\providecommand \BibitemOpen [0]{}%
\providecommand \bibitemStop [0]{}%
\providecommand \bibitemNoStop [0]{.\EOS\space}%
\providecommand \EOS [0]{\spacefactor3000\relax}%
\providecommand \BibitemShut  [1]{\csname bibitem#1\endcsname}%
\let\auto@bib@innerbib\@empty
\bibitem [{\citenamefont {Kohn}\ and\ \citenamefont {Sham}(1965)}]{KS65}%
  \BibitemOpen
  \bibfield  {author} {\bibinfo {author} {\bibfnamefont {W.}~\bibnamefont
  {Kohn}}\ and\ \bibinfo {author} {\bibfnamefont {L.~J.}\ \bibnamefont
  {Sham}},\ }\href {\doibase 10.1103/PhysRev.140.A1133} {\bibfield  {journal}
  {\bibinfo  {journal} {Phys. Rev.}\ }\textbf {\bibinfo {volume} {140}},\
  \bibinfo {pages} {A1133} (\bibinfo {year} {1965})}\BibitemShut {NoStop}%
\bibitem [{\citenamefont {Pribram-Jones}, \citenamefont {Gross},\ and\
  \citenamefont {Burke}(2015)}]{PGB15}%
  \BibitemOpen
  \bibfield  {author} {\bibinfo {author} {\bibfnamefont {A.}~\bibnamefont
  {Pribram-Jones}}, \bibinfo {author} {\bibfnamefont {D.~A.}\ \bibnamefont
  {Gross}}, \ and\ \bibinfo {author} {\bibfnamefont {K.}~\bibnamefont
  {Burke}},\ }\href {http://arxiv.org/abs/1408.4826} {\bibfield  {journal}
  {\bibinfo  {journal} {Annu. Rev. Phys. Chem.}\ } (\bibinfo {year}
  {2015})}\BibitemShut {NoStop}%
\bibitem [{\citenamefont {Fiolhais}, \citenamefont {Nogueira},\ and\
  \citenamefont {Marques}(2003)}]{FNM03}%
  \BibitemOpen
  \bibfield  {author} {\bibinfo {author} {\bibfnamefont {C.}~\bibnamefont
  {Fiolhais}}, \bibinfo {author} {\bibfnamefont {F.}~\bibnamefont {Nogueira}},
  \ and\ \bibinfo {author} {\bibfnamefont {M.}~\bibnamefont {Marques}},\ }\href
  {http://books.google.com/books?id=mX793GABep8C&dq=A+Primer+in+Density+Functional+Theory&printsec=frontcover&source=bn&hl=en&ei=IVt8TNDSEYeCsQOa0bCEBw&sa=X&oi=book_result&ct=result&resnum=4&ved=0CCYQ6AEwAw\#v=onepage&q&f=false}
  {\emph {\bibinfo {title} {A Primer in Density Functional Theory}}}\ (\bibinfo
   {publisher} {Springer-Verlag},\ \bibinfo {address} {New York},\ \bibinfo
  {year} {2003})\BibitemShut {NoStop}%
\bibitem [{\citenamefont {Becke}(1988)}]{B88}%
  \BibitemOpen
  \bibfield  {author} {\bibinfo {author} {\bibfnamefont {A.~D.}\ \bibnamefont
  {Becke}},\ }\href {\doibase 10.1103/PhysRevA.38.3098} {\bibfield  {journal}
  {\bibinfo  {journal} {Phys. Rev. A}\ }\textbf {\bibinfo {volume} {38}},\
  \bibinfo {pages} {3098} (\bibinfo {year} {1988})}\BibitemShut {NoStop}%
\bibitem [{\citenamefont {Lee}, \citenamefont {Yang},\ and\ \citenamefont
  {Parr}(1988)}]{LYP88}%
  \BibitemOpen
  \bibfield  {author} {\bibinfo {author} {\bibfnamefont {C.}~\bibnamefont
  {Lee}}, \bibinfo {author} {\bibfnamefont {W.}~\bibnamefont {Yang}}, \ and\
  \bibinfo {author} {\bibfnamefont {R.~G.}\ \bibnamefont {Parr}},\ }\href
  {\doibase 10.1103/PhysRevB.37.785} {\bibfield  {journal} {\bibinfo  {journal}
  {Phys. Rev. B}\ }\textbf {\bibinfo {volume} {37}},\ \bibinfo {pages} {785}
  (\bibinfo {year} {1988})}\BibitemShut {NoStop}%
\bibitem [{\citenamefont {Becke}(1993)}]{Bb93}%
  \BibitemOpen
  \bibfield  {author} {\bibinfo {author} {\bibfnamefont {A.~D.}\ \bibnamefont
  {Becke}},\ }\href {\doibase 10.1063/1.464913} {\bibfield  {journal} {\bibinfo
   {journal} {J. Chem. Phys.}\ }\textbf {\bibinfo {volume} {98}},\ \bibinfo
  {pages} {5648} (\bibinfo {year} {1993})}\BibitemShut {NoStop}%
\bibitem [{\citenamefont {Perdew}\ \emph {et~al.}(1992)\citenamefont {Perdew},
  \citenamefont {Chevary}, \citenamefont {Vosko}, \citenamefont {Jackson},
  \citenamefont {Pederson}, \citenamefont {Singh},\ and\ \citenamefont
  {Fiolhais}}]{PCVJ92}%
  \BibitemOpen
  \bibfield  {author} {\bibinfo {author} {\bibfnamefont {J.~P.}\ \bibnamefont
  {Perdew}}, \bibinfo {author} {\bibfnamefont {J.~A.}\ \bibnamefont {Chevary}},
  \bibinfo {author} {\bibfnamefont {S.~H.}\ \bibnamefont {Vosko}}, \bibinfo
  {author} {\bibfnamefont {K.~A.}\ \bibnamefont {Jackson}}, \bibinfo {author}
  {\bibfnamefont {M.~R.}\ \bibnamefont {Pederson}}, \bibinfo {author}
  {\bibfnamefont {D.~J.}\ \bibnamefont {Singh}}, \ and\ \bibinfo {author}
  {\bibfnamefont {C.}~\bibnamefont {Fiolhais}},\ }\href {\doibase
  10.1103/PhysRevB.46.6671} {\bibfield  {journal} {\bibinfo  {journal} {Phys.
  Rev. B}\ }\textbf {\bibinfo {volume} {46}},\ \bibinfo {pages} {6671}
  (\bibinfo {year} {1992})}\BibitemShut {NoStop}%
\bibitem [{\citenamefont {Perdew}, \citenamefont {Burke},\ and\ \citenamefont
  {Ernzerhof}(1996)}]{PBE96}%
  \BibitemOpen
  \bibfield  {author} {\bibinfo {author} {\bibfnamefont {J.~P.}\ \bibnamefont
  {Perdew}}, \bibinfo {author} {\bibfnamefont {K.}~\bibnamefont {Burke}}, \
  and\ \bibinfo {author} {\bibfnamefont {M.}~\bibnamefont {Ernzerhof}},\ }\href
  {\doibase 10.1103/PhysRevLett.77.3865} {\bibfield  {journal} {\bibinfo
  {journal} {Phys. Rev. Lett.}\ }\textbf {\bibinfo {volume} {77}},\ \bibinfo
  {pages} {3865} (\bibinfo {year} {1996})},\ \bibinfo {note} {{\it ibid.} {\bf
  78}, 1396(E) (1997)}\BibitemShut {NoStop}%
\bibitem [{\citenamefont {K\"ummel}\ and\ \citenamefont {Kronik}(2008)}]{KK08}%
  \BibitemOpen
  \bibfield  {author} {\bibinfo {author} {\bibfnamefont {S.}~\bibnamefont
  {K\"ummel}}\ and\ \bibinfo {author} {\bibfnamefont {L.}~\bibnamefont
  {Kronik}},\ }\href {\doibase 10.1103/RevModPhys.80.3} {\bibfield  {journal}
  {\bibinfo  {journal} {Rev. Mod. Phys.}\ }\textbf {\bibinfo {volume} {80}},\
  \bibinfo {pages} {3} (\bibinfo {year} {2008})}\BibitemShut {NoStop}%
\bibitem [{\citenamefont {Dion}\ \emph {et~al.}(2004)\citenamefont {Dion},
  \citenamefont {Rydberg}, \citenamefont {Schr\"oder}, \citenamefont
  {Langreth},\ and\ \citenamefont {Lundqvist}}]{DRSL04}%
  \BibitemOpen
  \bibfield  {author} {\bibinfo {author} {\bibfnamefont {M.}~\bibnamefont
  {Dion}}, \bibinfo {author} {\bibfnamefont {H.}~\bibnamefont {Rydberg}},
  \bibinfo {author} {\bibfnamefont {E.}~\bibnamefont {Schr\"oder}}, \bibinfo
  {author} {\bibfnamefont {D.~C.}\ \bibnamefont {Langreth}}, \ and\ \bibinfo
  {author} {\bibfnamefont {B.~I.}\ \bibnamefont {Lundqvist}},\ }\href {\doibase
  10.1103/PhysRevLett.92.246401} {\bibfield  {journal} {\bibinfo  {journal}
  {Phys. Rev. Lett.}\ }\textbf {\bibinfo {volume} {92}},\ \bibinfo {pages}
  {246401} (\bibinfo {year} {2004})}\BibitemShut {NoStop}%
\bibitem [{\citenamefont {Levy}(1979)}]{L79}%
  \BibitemOpen
  \bibfield  {author} {\bibinfo {author} {\bibfnamefont {M.}~\bibnamefont
  {Levy}},\ }\href {http://www.pnas.org/content/76/12/6062.abstract} {\bibfield
   {journal} {\bibinfo  {journal} {Proc. Natl. Acad. Sci. U. S. A.}\ }\textbf
  {\bibinfo {volume} {76}},\ \bibinfo {pages} {6062} (\bibinfo {year}
  {1979})}\BibitemShut {NoStop}%
\bibitem [{\citenamefont {Lieb}(1981)}]{L81}%
  \BibitemOpen
  \bibfield  {author} {\bibinfo {author} {\bibfnamefont {E.~H.}\ \bibnamefont
  {Lieb}},\ }\href {\doibase 10.1103/RevModPhys.53.603} {\bibfield  {journal}
  {\bibinfo  {journal} {Rev. Mod. Phys.}\ }\textbf {\bibinfo {volume} {53}},\
  \bibinfo {pages} {603} (\bibinfo {year} {1981})}\BibitemShut {NoStop}%
\bibitem [{\citenamefont {Lieb}\ and\ \citenamefont {Simon}(1977)}]{LS77}%
  \BibitemOpen
  \bibfield  {author} {\bibinfo {author} {\bibfnamefont {E.~H.}\ \bibnamefont
  {Lieb}}\ and\ \bibinfo {author} {\bibfnamefont {B.}~\bibnamefont {Simon}},\
  }\href@noop {} {\bibfield  {journal} {\bibinfo  {journal} {Advances in
  Mathematics}\ }\textbf {\bibinfo {volume} {23}},\ \bibinfo {pages} {22}
  (\bibinfo {year} {1977})}\BibitemShut {NoStop}%
\bibitem [{\citenamefont {Elliott}\ \emph {et~al.}(2008)\citenamefont
  {Elliott}, \citenamefont {Lee}, \citenamefont {Cangi},\ and\ \citenamefont
  {Burke}}]{ELCB08}%
  \BibitemOpen
  \bibfield  {author} {\bibinfo {author} {\bibfnamefont {P.}~\bibnamefont
  {Elliott}}, \bibinfo {author} {\bibfnamefont {D.}~\bibnamefont {Lee}},
  \bibinfo {author} {\bibfnamefont {A.}~\bibnamefont {Cangi}}, \ and\ \bibinfo
  {author} {\bibfnamefont {K.}~\bibnamefont {Burke}},\ }\href {\doibase
  10.1103/PhysRevLett.100.256406} {\bibfield  {journal} {\bibinfo  {journal}
  {Phys. Rev. Lett.}\ }\textbf {\bibinfo {volume} {100}},\ \bibinfo {pages}
  {256406} (\bibinfo {year} {2008})}\BibitemShut {NoStop}%
\bibitem [{\citenamefont {Elliott}\ \emph {et~al.}(2015)\citenamefont
  {Elliott}, \citenamefont {Cangi}, \citenamefont {Pittalis}, \citenamefont
  {Gross},\ and\ \citenamefont {Burke}}]{ECPG15}%
  \BibitemOpen
  \bibfield  {author} {\bibinfo {author} {\bibfnamefont {P.}~\bibnamefont
  {Elliott}}, \bibinfo {author} {\bibfnamefont {A.}~\bibnamefont {Cangi}},
  \bibinfo {author} {\bibfnamefont {S.}~\bibnamefont {Pittalis}}, \bibinfo
  {author} {\bibfnamefont {E.~K.~U.}\ \bibnamefont {Gross}}, \ and\ \bibinfo
  {author} {\bibfnamefont {K.}~\bibnamefont {Burke}},\ }\href {\doibase
  10.1103/PhysRevA.92.022513} {\bibfield  {journal} {\bibinfo  {journal} {Phys.
  Rev. A}\ }\textbf {\bibinfo {volume} {92}},\ \bibinfo {pages} {022513}
  (\bibinfo {year} {2015})}\BibitemShut {NoStop}%
\bibitem [{\citenamefont {Ribeiro}\ \emph {et~al.}(2015)\citenamefont
  {Ribeiro}, \citenamefont {Lee}, \citenamefont {Cangi}, \citenamefont
  {Elliott},\ and\ \citenamefont {Burke}}]{RLCE15}%
  \BibitemOpen
  \bibfield  {author} {\bibinfo {author} {\bibfnamefont {R.~F.}\ \bibnamefont
  {Ribeiro}}, \bibinfo {author} {\bibfnamefont {D.}~\bibnamefont {Lee}},
  \bibinfo {author} {\bibfnamefont {A.}~\bibnamefont {Cangi}}, \bibinfo
  {author} {\bibfnamefont {P.}~\bibnamefont {Elliott}}, \ and\ \bibinfo
  {author} {\bibfnamefont {K.}~\bibnamefont {Burke}},\ }\href {\doibase
  10.1103/PhysRevLett.114.050401} {\bibfield  {journal} {\bibinfo  {journal}
  {Phys. Rev. Lett.}\ }\textbf {\bibinfo {volume} {114}},\ \bibinfo {pages}
  {050401} (\bibinfo {year} {2015})}\BibitemShut {NoStop}%
\bibitem [{\citenamefont {Schwinger}(1980)}]{S80}%
  \BibitemOpen
  \bibfield  {author} {\bibinfo {author} {\bibfnamefont {J.}~\bibnamefont
  {Schwinger}},\ }\href@noop {} {\bibfield  {journal} {\bibinfo  {journal}
  {Phys. Rev. A}\ }\textbf {\bibinfo {volume} {22}},\ \bibinfo {pages} {1827}
  (\bibinfo {year} {1980})}\BibitemShut {NoStop}%
\bibitem [{\citenamefont {Schwinger}(1981)}]{S81}%
  \BibitemOpen
  \bibfield  {author} {\bibinfo {author} {\bibfnamefont {J.}~\bibnamefont
  {Schwinger}},\ }\href {\doibase 10.1103/PhysRevA.24.2353} {\bibfield
  {journal} {\bibinfo  {journal} {Phys. Rev. A}\ }\textbf {\bibinfo {volume}
  {24}},\ \bibinfo {pages} {2353} (\bibinfo {year} {1981})}\BibitemShut
  {NoStop}%
\bibitem [{\citenamefont {Conlon}(1983)}]{C83}%
  \BibitemOpen
  \bibfield  {author} {\bibinfo {author} {\bibfnamefont {J.}~\bibnamefont
  {Conlon}},\ }\href {\doibase 10.1007/BF01206884} {\bibfield  {journal}
  {\bibinfo  {journal} {Communications in Mathematical Physics}\ }\textbf
  {\bibinfo {volume} {88}},\ \bibinfo {pages} {133} (\bibinfo {year}
  {1983})}\BibitemShut {NoStop}%
\bibitem [{\citenamefont {Fefferman}\ and\ \citenamefont {Seco}(1994)}]{FSb94}%
  \BibitemOpen
  \bibfield  {author} {\bibinfo {author} {\bibfnamefont {C.}~\bibnamefont
  {Fefferman}}\ and\ \bibinfo {author} {\bibfnamefont {L.}~\bibnamefont
  {Seco}},\ }\href {\doibase 10.1006/aima.1994.1060} {\bibfield  {journal}
  {\bibinfo  {journal} {Advances in Mathematics}\ }\textbf {\bibinfo {volume}
  {107}},\ \bibinfo {pages} {1} (\bibinfo {year} {1994})}\BibitemShut {NoStop}%
\bibitem [{\citenamefont {Elliott}\ and\ \citenamefont {Burke}(2009)}]{EB09}%
  \BibitemOpen
  \bibfield  {author} {\bibinfo {author} {\bibfnamefont {P.}~\bibnamefont
  {Elliott}}\ and\ \bibinfo {author} {\bibfnamefont {K.}~\bibnamefont
  {Burke}},\ }\href {\doibase 10.1139/V09-095} {\bibfield  {journal} {\bibinfo
  {journal} {Can. J. Chem.}\ }\textbf {\bibinfo {volume} {87}},\ \bibinfo
  {pages} {1485} (\bibinfo {year} {2009})}\BibitemShut {NoStop}%
\bibitem [{\citenamefont {Perdew}\ \emph {et~al.}(2008)\citenamefont {Perdew},
  \citenamefont {Ruzsinszky}, \citenamefont {Csonka}, \citenamefont {Vydrov},
  \citenamefont {Scuseria}, \citenamefont {Constantin}, \citenamefont {Zhou},\
  and\ \citenamefont {Burke}}]{PRCV08}%
  \BibitemOpen
  \bibfield  {author} {\bibinfo {author} {\bibfnamefont {J.~P.}\ \bibnamefont
  {Perdew}}, \bibinfo {author} {\bibfnamefont {A.}~\bibnamefont {Ruzsinszky}},
  \bibinfo {author} {\bibfnamefont {G.~I.}\ \bibnamefont {Csonka}}, \bibinfo
  {author} {\bibfnamefont {O.~A.}\ \bibnamefont {Vydrov}}, \bibinfo {author}
  {\bibfnamefont {G.~E.}\ \bibnamefont {Scuseria}}, \bibinfo {author}
  {\bibfnamefont {L.~A.}\ \bibnamefont {Constantin}}, \bibinfo {author}
  {\bibfnamefont {X.}~\bibnamefont {Zhou}}, \ and\ \bibinfo {author}
  {\bibfnamefont {K.}~\bibnamefont {Burke}},\ }\href {\doibase
  10.1103/PhysRevLett.100.136406} {\bibfield  {journal} {\bibinfo  {journal}
  {Phys. Rev. Lett.}\ }\textbf {\bibinfo {volume} {100}},\ \bibinfo {pages}
  {136406} (\bibinfo {year} {2008})}\BibitemShut {NoStop}%
\bibitem [{\citenamefont {Zhang}\ and\ \citenamefont {Yang}(1998)}]{ZY98}%
  \BibitemOpen
  \bibfield  {author} {\bibinfo {author} {\bibfnamefont {Y.}~\bibnamefont
  {Zhang}}\ and\ \bibinfo {author} {\bibfnamefont {W.}~\bibnamefont {Yang}},\
  }\href@noop {} {\bibfield  {journal} {\bibinfo  {journal} {Phys. Rev. Lett.}\
  }\textbf {\bibinfo {volume} {80}},\ \bibinfo {pages} {890} (\bibinfo {year}
  {1998})}\BibitemShut {NoStop}%
\bibitem [{\citenamefont {Perdew}, \citenamefont {Burke},\ and\ \citenamefont
  {Ernzerhof}(1998)}]{PBE98}%
  \BibitemOpen
  \bibfield  {author} {\bibinfo {author} {\bibfnamefont {J.}~\bibnamefont
  {Perdew}}, \bibinfo {author} {\bibfnamefont {K.}~\bibnamefont {Burke}}, \
  and\ \bibinfo {author} {\bibfnamefont {M.}~\bibnamefont {Ernzerhof}},\
  }\href@noop {} {\bibfield  {journal} {\bibinfo  {journal} {Phys. Rev. Lett.}\
  }\textbf {\bibinfo {volume} {80}},\ \bibinfo {pages} {891} (\bibinfo {year}
  {1998})}\BibitemShut {NoStop}%
\bibitem [{\citenamefont {Constantin}\ \emph {et~al.}(2016)\citenamefont
  {Constantin}, \citenamefont {Terentjevs}, \citenamefont {Della~Sala},
  \citenamefont {Cortona},\ and\ \citenamefont {Fabiano}}]{CTDC16}%
  \BibitemOpen
  \bibfield  {author} {\bibinfo {author} {\bibfnamefont {L.~A.}\ \bibnamefont
  {Constantin}}, \bibinfo {author} {\bibfnamefont {A.}~\bibnamefont
  {Terentjevs}}, \bibinfo {author} {\bibfnamefont {F.}~\bibnamefont
  {Della~Sala}}, \bibinfo {author} {\bibfnamefont {P.}~\bibnamefont {Cortona}},
  \ and\ \bibinfo {author} {\bibfnamefont {E.}~\bibnamefont {Fabiano}},\ }\href
  {\doibase 10.1103/PhysRevB.93.045126} {\bibfield  {journal} {\bibinfo
  {journal} {Phys. Rev. B}\ }\textbf {\bibinfo {volume} {93}},\ \bibinfo
  {pages} {045126} (\bibinfo {year} {2016})}\BibitemShut {NoStop}%
\bibitem [{\citenamefont {Laricchia}\ \emph {et~al.}(2011)\citenamefont
  {Laricchia}, \citenamefont {Fabiano}, \citenamefont {Constantin},\ and\
  \citenamefont {Sala}}]{LFCD11}%
  \BibitemOpen
  \bibfield  {author} {\bibinfo {author} {\bibfnamefont {S.}~\bibnamefont
  {Laricchia}}, \bibinfo {author} {\bibfnamefont {E.}~\bibnamefont {Fabiano}},
  \bibinfo {author} {\bibfnamefont {L.~A.}\ \bibnamefont {Constantin}}, \ and\
  \bibinfo {author} {\bibfnamefont {F.~D.}\ \bibnamefont {Sala}},\ }\href
  {\doibase 10.1021/ct200382w} {\bibfield  {journal} {\bibinfo  {journal} {J.
  Chem. Theory Comput.}\ }\textbf {\bibinfo {volume} {7}},\ \bibinfo {pages}
  {2439} (\bibinfo {year} {2011})},\ \bibinfo {note} {pMID: 26606618},\ \Eprint
  {http://arxiv.org/abs/http://dx.doi.org/10.1021/ct200382w}
  {http://dx.doi.org/10.1021/ct200382w} \BibitemShut {NoStop}%
\bibitem [{\citenamefont {Constantin}\ \emph {et~al.}(2011)\citenamefont
  {Constantin}, \citenamefont {Fabiano}, \citenamefont {Laricchia},\ and\
  \citenamefont {Della~Sala}}]{CFLD11}%
  \BibitemOpen
  \bibfield  {author} {\bibinfo {author} {\bibfnamefont {L.~A.}\ \bibnamefont
  {Constantin}}, \bibinfo {author} {\bibfnamefont {E.}~\bibnamefont {Fabiano}},
  \bibinfo {author} {\bibfnamefont {S.}~\bibnamefont {Laricchia}}, \ and\
  \bibinfo {author} {\bibfnamefont {F.}~\bibnamefont {Della~Sala}},\ }\href
  {\doibase 10.1103/PhysRevLett.106.186406} {\bibfield  {journal} {\bibinfo
  {journal} {Phys. Rev. Lett.}\ }\textbf {\bibinfo {volume} {106}},\ \bibinfo
  {pages} {186406} (\bibinfo {year} {2011})}\BibitemShut {NoStop}%
\bibitem [{\citenamefont {McCarthy}\ and\ \citenamefont
  {Thakkar}(2011)}]{MT11}%
  \BibitemOpen
  \bibfield  {author} {\bibinfo {author} {\bibfnamefont {S.~P.}\ \bibnamefont
  {McCarthy}}\ and\ \bibinfo {author} {\bibfnamefont {A.~J.}\ \bibnamefont
  {Thakkar}},\ }\href {\doibase http://dx.doi.org/10.1063/1.3547262} {\bibfield
   {journal} {\bibinfo  {journal} {J. Chem. Phys.}\ }\textbf {\bibinfo {volume}
  {134}},\ \bibinfo {eid} {044102} (\bibinfo {year} {2011})}\BibitemShut
  {NoStop}%
\bibitem [{\citenamefont {Kunz}\ and\ \citenamefont {Rueedi}(2010)}]{KR10}%
  \BibitemOpen
  \bibfield  {author} {\bibinfo {author} {\bibfnamefont {H.}~\bibnamefont
  {Kunz}}\ and\ \bibinfo {author} {\bibfnamefont {R.}~\bibnamefont {Rueedi}},\
  }\href {\doibase 10.1103/PhysRevA.81.032122} {\bibfield  {journal} {\bibinfo
  {journal} {Phys. Rev. A}\ }\textbf {\bibinfo {volume} {81}},\ \bibinfo
  {pages} {032122} (\bibinfo {year} {2010})}\BibitemShut {NoStop}%
\bibitem [{\citenamefont {Clementi}\ and\ \citenamefont
  {Corongiu}(1997)}]{Clementi1997}%
  \BibitemOpen
  \bibfield  {author} {\bibinfo {author} {\bibfnamefont {E.}~\bibnamefont
  {Clementi}}\ and\ \bibinfo {author} {\bibfnamefont {G.}~\bibnamefont
  {Corongiu}},\ }\href {\doibase
  10.1002/(SICI)1097-461X(1997)62:6<571::AID-QUA2>3.0.CO;2-T} {\bibfield
  {journal} {\bibinfo  {journal} {Int. J. Quantum Chem.}\ }\textbf {\bibinfo
  {volume} {62}},\ \bibinfo {pages} {571} (\bibinfo {year} {1997})}\BibitemShut
  {NoStop}%
\bibitem [{foo({\natexlab{a}})}]{footnote1}%
  \BibitemOpen
  \bibinfo {note} {The values reported for
  $A\c$ and $B\c$ on pages 9 and 10 of Ref.~\onlinecite{KR10} appear to have
  typographical errors. To reproduce Fig. 1, we divide their $A\c$ by three and
  reverse the sign of $B\c$. But even with these corrections, the resulting
  value for $B\c$ is inaccurate.}\BibitemShut {Stop}%
\bibitem [{\citenamefont {Englert}(1988)}]{E88}%
  \BibitemOpen
  \bibfield  {author} {\bibinfo {author} {\bibfnamefont {B.-G.}\ \bibnamefont
  {Englert}},\ }\href@noop {} {\emph {\bibinfo {title} {Semiclassical theory of
  atoms}}},\ \bibinfo {series} {Lecture Notes in Physics}, Vol.\ \bibinfo
  {volume} {300}\ (\bibinfo  {publisher} {Springer Verlag, Berlin},\ \bibinfo
  {year} {1988})\BibitemShut {NoStop}%
\bibitem [{\citenamefont {Hohenberg}\ and\ \citenamefont {Kohn}(1964)}]{HK64}%
  \BibitemOpen
  \bibfield  {author} {\bibinfo {author} {\bibfnamefont {P.}~\bibnamefont
  {Hohenberg}}\ and\ \bibinfo {author} {\bibfnamefont {W.}~\bibnamefont
  {Kohn}},\ }\href {\doibase 10.1103/PhysRev.136.B864} {\bibfield  {journal}
  {\bibinfo  {journal} {Phys. Rev.}\ }\textbf {\bibinfo {volume} {136}},\
  \bibinfo {pages} {B864} (\bibinfo {year} {1964})}\BibitemShut {NoStop}%
\bibitem [{\citenamefont {Thomas}(1927)}]{T27}%
  \BibitemOpen
  \bibfield  {author} {\bibinfo {author} {\bibfnamefont {L.~H.}\ \bibnamefont
  {Thomas}},\ }\href {\doibase 10.1017/S0305004100011683} {\bibfield  {journal}
  {\bibinfo  {journal} {Math. Proc. Camb. Phil. Soc.}\ }\textbf {\bibinfo
  {volume} {23}},\ \bibinfo {pages} {542} (\bibinfo {year} {1927})}\BibitemShut
  {NoStop}%
\bibitem [{\citenamefont {Fermi}(1928)}]{F28}%
  \BibitemOpen
  \bibfield  {author} {\bibinfo {author} {\bibfnamefont {E.}~\bibnamefont
  {Fermi}},\ }\href {\doibase 10.1007/BF01351576} {\bibfield  {journal}
  {\bibinfo  {journal} {Zeitschrift f\"ur Physik A Hadrons and Nuclei}\
  }\textbf {\bibinfo {volume} {48}},\ \bibinfo {pages} {73} (\bibinfo {year}
  {1928})}\BibitemShut {NoStop}%
\bibitem [{\citenamefont {Spruch}(1991)}]{Sa91}%
  \BibitemOpen
  \bibfield  {author} {\bibinfo {author} {\bibfnamefont {L.}~\bibnamefont
  {Spruch}},\ }\href@noop {} {\bibfield  {journal} {\bibinfo  {journal} {Rev.
  Mod. Phys.}\ }\textbf {\bibinfo {volume} {63}},\ \bibinfo {pages} {151}
  (\bibinfo {year} {1991})}\BibitemShut {NoStop}%
\bibitem [{\citenamefont {Lee}\ \emph {et~al.}(2009)\citenamefont {Lee},
  \citenamefont {Constantin}, \citenamefont {Perdew},\ and\ \citenamefont
  {Burke}}]{LCPB09}%
  \BibitemOpen
  \bibfield  {author} {\bibinfo {author} {\bibfnamefont {D.}~\bibnamefont
  {Lee}}, \bibinfo {author} {\bibfnamefont {L.~A.}\ \bibnamefont {Constantin}},
  \bibinfo {author} {\bibfnamefont {J.~P.}\ \bibnamefont {Perdew}}, \ and\
  \bibinfo {author} {\bibfnamefont {K.}~\bibnamefont {Burke}},\ }\href
  {\doibase 10.1063/1.3059783} {\bibfield  {journal} {\bibinfo  {journal} {J.
  Chem. Phys.}\ }\textbf {\bibinfo {volume} {130}},\ \bibinfo {eid} {034107}
  (\bibinfo {year} {2009})}\BibitemShut {NoStop}%
\bibitem [{\citenamefont {von Barth}\ and\ \citenamefont {Hedin}(1972)}]{BH72}%
  \BibitemOpen
  \bibfield  {author} {\bibinfo {author} {\bibfnamefont {U.}~\bibnamefont {von
  Barth}}\ and\ \bibinfo {author} {\bibfnamefont {L.}~\bibnamefont {Hedin}},\
  }\href {http://stacks.iop.org/0022-3719/5/i=13/a=012} {\bibfield  {journal}
  {\bibinfo  {journal} {Journal of Physics C: Solid State Physics}\ }\textbf
  {\bibinfo {volume} {5}},\ \bibinfo {pages} {1629} (\bibinfo {year}
  {1972})}\BibitemShut {NoStop}%
\bibitem [{\citenamefont {Ceperley}\ and\ \citenamefont {Alder}(1980)}]{CA80}%
  \BibitemOpen
  \bibfield  {author} {\bibinfo {author} {\bibfnamefont {D.~M.}\ \bibnamefont
  {Ceperley}}\ and\ \bibinfo {author} {\bibfnamefont {B.~J.}\ \bibnamefont
  {Alder}},\ }\href@noop {} {\bibfield  {journal} {\bibinfo  {journal} {Phys.
  Rev. Lett.}\ }\textbf {\bibinfo {volume} {45}},\ \bibinfo {pages} {566}
  (\bibinfo {year} {1980})}\BibitemShut {NoStop}%
\bibitem [{\citenamefont {Vosko}, \citenamefont {Wilk},\ and\ \citenamefont
  {Nusair}(1980)}]{VWN80}%
  \BibitemOpen
  \bibfield  {author} {\bibinfo {author} {\bibfnamefont {S.~H.}\ \bibnamefont
  {Vosko}}, \bibinfo {author} {\bibfnamefont {L.}~\bibnamefont {Wilk}}, \ and\
  \bibinfo {author} {\bibfnamefont {M.}~\bibnamefont {Nusair}},\ }\href
  {\doibase 10.1139/p80-159} {\bibfield  {journal} {\bibinfo  {journal} {Can.
  J. Phys.}\ }\textbf {\bibinfo {volume} {58}},\ \bibinfo {pages} {1200}
  (\bibinfo {year} {1980})},\ \Eprint
  {http://arxiv.org/abs/http://dx.doi.org/10.1139/p80-159}
  {http://dx.doi.org/10.1139/p80-159} \BibitemShut {NoStop}%
\bibitem [{\citenamefont {Perdew}\ and\ \citenamefont {Wang}(1992)}]{PW92}%
  \BibitemOpen
  \bibfield  {author} {\bibinfo {author} {\bibfnamefont {J.~P.}\ \bibnamefont
  {Perdew}}\ and\ \bibinfo {author} {\bibfnamefont {Y.}~\bibnamefont {Wang}},\
  }\href {\doibase 10.1103/PhysRevB.45.13244} {\bibfield  {journal} {\bibinfo
  {journal} {Phys. Rev. B}\ }\textbf {\bibinfo {volume} {45}},\ \bibinfo
  {pages} {13244} (\bibinfo {year} {1992})}\BibitemShut {NoStop}%
\bibitem [{\citenamefont {Jones}\ and\ \citenamefont
  {Gunnarsson}(1989)}]{JG89}%
  \BibitemOpen
  \bibfield  {author} {\bibinfo {author} {\bibfnamefont {R.}~\bibnamefont
  {Jones}}\ and\ \bibinfo {author} {\bibfnamefont {O.}~\bibnamefont
  {Gunnarsson}},\ }\href@noop {} {\bibfield  {journal} {\bibinfo  {journal}
  {Rev. Mod. Phys.}\ }\textbf {\bibinfo {volume} {61}},\ \bibinfo {pages} {689}
  (\bibinfo {year} {1989})}\BibitemShut {NoStop}%
\bibitem [{\citenamefont {Jones}(2015)}]{J15}%
  \BibitemOpen
  \bibfield  {author} {\bibinfo {author} {\bibfnamefont {R.~O.}\ \bibnamefont
  {Jones}},\ }\href {\doibase 10.1103/RevModPhys.87.897} {\bibfield  {journal}
  {\bibinfo  {journal} {Rev. Mod. Phys.}\ }\textbf {\bibinfo {volume} {87}},\
  \bibinfo {pages} {897} (\bibinfo {year} {2015})}\BibitemShut {NoStop}%
\bibitem [{\citenamefont {Dreizler}\ and\ \citenamefont {Gross}(1990)}]{DG90}%
  \BibitemOpen
  \bibfield  {author} {\bibinfo {author} {\bibfnamefont {R.~M.}\ \bibnamefont
  {Dreizler}}\ and\ \bibinfo {author} {\bibfnamefont {E.~K.~U.}\ \bibnamefont
  {Gross}},\ }\href@noop {} {\emph {\bibinfo {title} {Density Functional
  Theory: An Approach to the Quantum Many-Body Problem}}}\ (\bibinfo
  {publisher} {Springer--Verlag},\ \bibinfo {address} {Berlin},\ \bibinfo
  {year} {1990})\BibitemShut {NoStop}%
\bibitem [{\citenamefont {Kleinman}\ and\ \citenamefont {Lee}(1988)}]{KL88}%
  \BibitemOpen
  \bibfield  {author} {\bibinfo {author} {\bibfnamefont {L.}~\bibnamefont
  {Kleinman}}\ and\ \bibinfo {author} {\bibfnamefont {S.}~\bibnamefont {Lee}},\
  }\href@noop {} {\bibfield  {journal} {\bibinfo  {journal} {Phys. Rev. B}\
  }\textbf {\bibinfo {volume} {37}},\ \bibinfo {pages} {4634} (\bibinfo {year}
  {1988})}\BibitemShut {NoStop}%
\bibitem [{\citenamefont {Ma}\ and\ \citenamefont {Brueckner}(1968)}]{MB68}%
  \BibitemOpen
  \bibfield  {author} {\bibinfo {author} {\bibfnamefont {S.-K.}\ \bibnamefont
  {Ma}}\ and\ \bibinfo {author} {\bibfnamefont {K.}~\bibnamefont {Brueckner}},\
  }\href@noop {} {\bibfield  {journal} {\bibinfo  {journal} {Phys. Rev.}\
  }\textbf {\bibinfo {volume} {165}},\ \bibinfo {pages} {18} (\bibinfo {year}
  {1968})}\BibitemShut {NoStop}%
\bibitem [{\citenamefont {Langreth}\ and\ \citenamefont {Perdew}(1977)}]{LP77}%
  \BibitemOpen
  \bibfield  {author} {\bibinfo {author} {\bibfnamefont {D.}~\bibnamefont
  {Langreth}}\ and\ \bibinfo {author} {\bibfnamefont {J.}~\bibnamefont
  {Perdew}},\ }\href@noop {} {\bibfield  {journal} {\bibinfo  {journal} {Phys.
  Rev. B}\ }\textbf {\bibinfo {volume} {15}},\ \bibinfo {pages} {2884}
  (\bibinfo {year} {1977})}\BibitemShut {NoStop}%
\bibitem [{\citenamefont {Becke}(2014)}]{B14}%
  \BibitemOpen
  \bibfield  {author} {\bibinfo {author} {\bibfnamefont {A.~D.}\ \bibnamefont
  {Becke}},\ }\href {\doibase http://dx.doi.org/10.1063/1.4869598} {\bibfield
  {journal} {\bibinfo  {journal} {{The Journal of Chemical Physics}}\ }\textbf
  {\bibinfo {volume} {140}},\ \bibinfo {eid} {18A301} (\bibinfo {year}
  {2014}),\ http://dx.doi.org/10.1063/1.4869598}\BibitemShut {NoStop}%
\bibitem [{\citenamefont {Engel}\ and\ \citenamefont {Dreizler}(2011)}]{ED11}%
  \BibitemOpen
  \bibfield  {author} {\bibinfo {author} {\bibfnamefont {E.}~\bibnamefont
  {Engel}}\ and\ \bibinfo {author} {\bibfnamefont {R.~M.}\ \bibnamefont
  {Dreizler}},\ }\href@noop {} {\emph {\bibinfo {title} {Density Functional
  Theory: An Advanced Course}}}\ (\bibinfo  {publisher} {Springer},\ \bibinfo
  {address} {Berlin},\ \bibinfo {year} {2011})\BibitemShut {NoStop}%
\bibitem [{\citenamefont {G\"{o}rling}\ and\ \citenamefont
  {Ernzerhof}(1995)}]{GE95}%
  \BibitemOpen
  \bibfield  {author} {\bibinfo {author} {\bibfnamefont {A.}~\bibnamefont
  {G\"{o}rling}}\ and\ \bibinfo {author} {\bibfnamefont {M.}~\bibnamefont
  {Ernzerhof}},\ }\href@noop {} {\bibfield  {journal} {\bibinfo  {journal}
  {Phys. Rev. A}\ }\textbf {\bibinfo {volume} {51}},\ \bibinfo {pages} {4501}
  (\bibinfo {year} {1995})}\BibitemShut {NoStop}%
\bibitem [{\citenamefont {Levy}\ and\ \citenamefont {Perdew}(1985)}]{LP85}%
  \BibitemOpen
  \bibfield  {author} {\bibinfo {author} {\bibfnamefont {M.}~\bibnamefont
  {Levy}}\ and\ \bibinfo {author} {\bibfnamefont {J.}~\bibnamefont {Perdew}},\
  }\href {\doibase 10.1103/PhysRevA.32.2010} {\bibfield  {journal} {\bibinfo
  {journal} {Phys. Rev. A}\ }\textbf {\bibinfo {volume} {32}},\ \bibinfo
  {pages} {2010} (\bibinfo {year} {1985})}\BibitemShut {NoStop}%
\bibitem [{\citenamefont {Harris}\ and\ \citenamefont {Jones}(1974)}]{HJ74}%
  \BibitemOpen
  \bibfield  {author} {\bibinfo {author} {\bibfnamefont {J.}~\bibnamefont
  {Harris}}\ and\ \bibinfo {author} {\bibfnamefont {R.}~\bibnamefont {Jones}},\
  }\href@noop {} {\bibfield  {journal} {\bibinfo  {journal} {J. Phys. F}\
  }\textbf {\bibinfo {volume} {4}},\ \bibinfo {pages} {1170} (\bibinfo {year}
  {1974})}\BibitemShut {NoStop}%
\bibitem [{\citenamefont {Langreth}\ and\ \citenamefont {Perdew}(1975)}]{LP75}%
  \BibitemOpen
  \bibfield  {author} {\bibinfo {author} {\bibfnamefont {D.}~\bibnamefont
  {Langreth}}\ and\ \bibinfo {author} {\bibfnamefont {J.}~\bibnamefont
  {Perdew}},\ }\href@noop {} {\bibfield  {journal} {\bibinfo  {journal} {Solid
  State Commun.}\ }\textbf {\bibinfo {volume} {17}},\ \bibinfo {pages} {1425}
  (\bibinfo {year} {1975})}\BibitemShut {NoStop}%
\bibitem [{\citenamefont {Gunnarsson}\ and\ \citenamefont
  {Lundqvist}(1976)}]{GL76}%
  \BibitemOpen
  \bibfield  {author} {\bibinfo {author} {\bibfnamefont {O.}~\bibnamefont
  {Gunnarsson}}\ and\ \bibinfo {author} {\bibfnamefont {B.}~\bibnamefont
  {Lundqvist}},\ }\href@noop {} {\bibfield  {journal} {\bibinfo  {journal}
  {Phys. Rev. B}\ }\textbf {\bibinfo {volume} {13}},\ \bibinfo {pages} {4274}
  (\bibinfo {year} {1976})}\BibitemShut {NoStop}%
\bibitem [{\citenamefont {Lieb}\ and\ \citenamefont {Simon}(1973)}]{LS73}%
  \BibitemOpen
  \bibfield  {author} {\bibinfo {author} {\bibfnamefont {E.}~\bibnamefont
  {Lieb}}\ and\ \bibinfo {author} {\bibfnamefont {B.}~\bibnamefont {Simon}},\
  }\href@noop {} {\bibfield  {journal} {\bibinfo  {journal} {Phys. Rev. Lett.}\
  }\textbf {\bibinfo {volume} {31}},\ \bibinfo {pages} {681} (\bibinfo {year}
  {1973})}\BibitemShut {NoStop}%
\bibitem [{\citenamefont {{Fournais}}, \citenamefont {{Lewin}},\ and\
  \citenamefont {{Solovej}}(2015)}]{FLS15}%
  \BibitemOpen
  \bibfield  {author} {\bibinfo {author} {\bibfnamefont {S.}~\bibnamefont
  {{Fournais}}}, \bibinfo {author} {\bibfnamefont {M.}~\bibnamefont {{Lewin}}},
  \ and\ \bibinfo {author} {\bibfnamefont {J.~P.}\ \bibnamefont {{Solovej}}},\
  }\href@noop {} {\bibfield  {journal} {\bibinfo  {journal} {ArXiv e-prints}\ }
  (\bibinfo {year} {2015})},\ \Eprint {http://arxiv.org/abs/1510.01124}
  {arXiv:1510.01124 [math-ph]} \BibitemShut {NoStop}%
\bibitem [{\citenamefont {Heilmann}\ and\ \citenamefont {Lieb}(1995)}]{HL95}%
  \BibitemOpen
  \bibfield  {author} {\bibinfo {author} {\bibfnamefont {O.~J.}\ \bibnamefont
  {Heilmann}}\ and\ \bibinfo {author} {\bibfnamefont {E.~H.}\ \bibnamefont
  {Lieb}},\ }\href {\doibase 10.1103/PhysRevA.52.3628} {\bibfield  {journal}
  {\bibinfo  {journal} {Phys. Rev. A}\ }\textbf {\bibinfo {volume} {52}},\
  \bibinfo {pages} {3628} (\bibinfo {year} {1995})}\BibitemShut {NoStop}%
\bibitem [{\citenamefont {Scott}(1952)}]{S52}%
  \BibitemOpen
  \bibfield  {author} {\bibinfo {author} {\bibfnamefont {J.}~\bibnamefont
  {Scott}},\ }\href@noop {} {\bibfield  {journal} {\bibinfo  {journal} {Philos.
  Mag.}\ }\textbf {\bibinfo {volume} {43}},\ \bibinfo {pages} {859} (\bibinfo
  {year} {1952})}\BibitemShut {NoStop}%
\bibitem [{\citenamefont {Levine}\ and\ \citenamefont {Soven}(1983)}]{LS83}%
  \BibitemOpen
  \bibfield  {author} {\bibinfo {author} {\bibfnamefont {Z.}~\bibnamefont
  {Levine}}\ and\ \bibinfo {author} {\bibfnamefont {P.}~\bibnamefont {Soven}},\
  }\href@noop {} {\bibfield  {journal} {\bibinfo  {journal} {Phys. Rev. Lett.}\
  }\textbf {\bibinfo {volume} {50}},\ \bibinfo {pages} {2074} (\bibinfo {year}
  {1983})}\BibitemShut {NoStop}%
\bibitem [{\citenamefont {Englert}\ and\ \citenamefont
  {Schwinger}(1982)}]{ES82}%
  \BibitemOpen
  \bibfield  {author} {\bibinfo {author} {\bibfnamefont {B.-G.}\ \bibnamefont
  {Englert}}\ and\ \bibinfo {author} {\bibfnamefont {J.}~\bibnamefont
  {Schwinger}},\ }\href {\doibase 10.1103/PhysRevA.26.2322} {\bibfield
  {journal} {\bibinfo  {journal} {Phys. Rev. A}\ }\textbf {\bibinfo {volume}
  {26}},\ \bibinfo {pages} {2322} (\bibinfo {year} {1982})}\BibitemShut
  {NoStop}%
\bibitem [{\citenamefont {Englert}\ and\ \citenamefont
  {Schwinger}(1984{\natexlab{a}})}]{ES84}%
  \BibitemOpen
  \bibfield  {author} {\bibinfo {author} {\bibfnamefont {B.-G.}\ \bibnamefont
  {Englert}}\ and\ \bibinfo {author} {\bibfnamefont {J.}~\bibnamefont
  {Schwinger}},\ }\href {\doibase 10.1103/PhysRevA.29.2339} {\bibfield
  {journal} {\bibinfo  {journal} {Phys. Rev. A}\ }\textbf {\bibinfo {volume}
  {29}},\ \bibinfo {pages} {2339} (\bibinfo {year}
  {1984}{\natexlab{a}})}\BibitemShut {NoStop}%
\bibitem [{\citenamefont {Englert}\ and\ \citenamefont
  {Schwinger}(1984{\natexlab{b}})}]{ESb84}%
  \BibitemOpen
  \bibfield  {author} {\bibinfo {author} {\bibfnamefont {B.-G.}\ \bibnamefont
  {Englert}}\ and\ \bibinfo {author} {\bibfnamefont {J.}~\bibnamefont
  {Schwinger}},\ }\href {\doibase 10.1103/PhysRevA.29.2353} {\bibfield
  {journal} {\bibinfo  {journal} {Phys. Rev. A}\ }\textbf {\bibinfo {volume}
  {29}},\ \bibinfo {pages} {2353} (\bibinfo {year}
  {1984}{\natexlab{b}})}\BibitemShut {NoStop}%
\bibitem [{\citenamefont {Englert}\ and\ \citenamefont
  {Schwinger}(1984{\natexlab{c}})}]{ESc84}%
  \BibitemOpen
  \bibfield  {author} {\bibinfo {author} {\bibfnamefont {B.-G.}\ \bibnamefont
  {Englert}}\ and\ \bibinfo {author} {\bibfnamefont {J.}~\bibnamefont
  {Schwinger}},\ }\href {\doibase 10.1103/PhysRevA.29.2331} {\bibfield
  {journal} {\bibinfo  {journal} {Phys. Rev. A}\ }\textbf {\bibinfo {volume}
  {29}},\ \bibinfo {pages} {2331} (\bibinfo {year}
  {1984}{\natexlab{c}})}\BibitemShut {NoStop}%
\bibitem [{\citenamefont {Englert}\ and\ \citenamefont
  {Schwinger}(1985{\natexlab{a}})}]{ES85}%
  \BibitemOpen
  \bibfield  {author} {\bibinfo {author} {\bibfnamefont {B.-G.}\ \bibnamefont
  {Englert}}\ and\ \bibinfo {author} {\bibfnamefont {J.}~\bibnamefont
  {Schwinger}},\ }\href@noop {} {\bibfield  {journal} {\bibinfo  {journal}
  {Phys. Rev. A}\ }\textbf {\bibinfo {volume} {32}},\ \bibinfo {pages} {26}
  (\bibinfo {year} {1985}{\natexlab{a}})}\BibitemShut {NoStop}%
\bibitem [{\citenamefont {Englert}\ and\ \citenamefont
  {Schwinger}(1985{\natexlab{b}})}]{ESb85}%
  \BibitemOpen
  \bibfield  {author} {\bibinfo {author} {\bibfnamefont {B.-G.}\ \bibnamefont
  {Englert}}\ and\ \bibinfo {author} {\bibfnamefont {J.}~\bibnamefont
  {Schwinger}},\ }\href@noop {} {\bibfield  {journal} {\bibinfo  {journal}
  {Phys. Rev. A}\ }\textbf {\bibinfo {volume} {32}},\ \bibinfo {pages} {36}
  (\bibinfo {year} {1985}{\natexlab{b}})}\BibitemShut {NoStop}%
\bibitem [{\citenamefont {Englert}\ and\ \citenamefont
  {Schwinger}(1985{\natexlab{c}})}]{ESc85}%
  \BibitemOpen
  \bibfield  {author} {\bibinfo {author} {\bibfnamefont {B.-G.}\ \bibnamefont
  {Englert}}\ and\ \bibinfo {author} {\bibfnamefont {J.}~\bibnamefont
  {Schwinger}},\ }\href@noop {} {\bibfield  {journal} {\bibinfo  {journal}
  {Phys. Rev. A}\ }\textbf {\bibinfo {volume} {32}},\ \bibinfo {pages} {47}
  (\bibinfo {year} {1985}{\natexlab{c}})}\BibitemShut {NoStop}%
\bibitem [{\citenamefont {Perdew}\ \emph {et~al.}(2006)\citenamefont {Perdew},
  \citenamefont {Constantin}, \citenamefont {Sagvolden},\ and\ \citenamefont
  {Burke}}]{PCSB06}%
  \BibitemOpen
  \bibfield  {author} {\bibinfo {author} {\bibfnamefont {J.~P.}\ \bibnamefont
  {Perdew}}, \bibinfo {author} {\bibfnamefont {L.~A.}\ \bibnamefont
  {Constantin}}, \bibinfo {author} {\bibfnamefont {E.}~\bibnamefont
  {Sagvolden}}, \ and\ \bibinfo {author} {\bibfnamefont {K.}~\bibnamefont
  {Burke}},\ }\href {\doibase 10.1103/PhysRevLett.97.223002} {\bibfield
  {journal} {\bibinfo  {journal} {Phys. Rev. Lett.}\ }\textbf {\bibinfo
  {volume} {97}},\ \bibinfo {pages} {223002} (\bibinfo {year}
  {2006})}\BibitemShut {NoStop}%
\bibitem [{\citenamefont {Burke}, \citenamefont {Ernzerhof},\ and\
  \citenamefont {Perdew}(1997)}]{BEP97}%
  \BibitemOpen
  \bibfield  {author} {\bibinfo {author} {\bibfnamefont {K.}~\bibnamefont
  {Burke}}, \bibinfo {author} {\bibfnamefont {M.}~\bibnamefont {Ernzerhof}}, \
  and\ \bibinfo {author} {\bibfnamefont {J.~P.}\ \bibnamefont {Perdew}},\
  }\href {\doibase DOI: 10.1016/S0009-2614(96)01373-5} {\bibfield  {journal}
  {\bibinfo  {journal} {Chem. Phys. Lett.}\ }\textbf {\bibinfo {volume}
  {265}},\ \bibinfo {pages} {115} (\bibinfo {year} {1997})}\BibitemShut
  {NoStop}%
\bibitem [{\citenamefont {McCarthy}\ and\ \citenamefont
  {Thakkar}(2012)}]{MT12}%
  \BibitemOpen
  \bibfield  {author} {\bibinfo {author} {\bibfnamefont {S.~P.}\ \bibnamefont
  {McCarthy}}\ and\ \bibinfo {author} {\bibfnamefont {A.~J.}\ \bibnamefont
  {Thakkar}},\ }\href {\doibase http://dx.doi.org/10.1063/1.3679969} {\bibfield
   {journal} {\bibinfo  {journal} {J. Chem. Phys.}\ }\textbf {\bibinfo {volume}
  {136}},\ \bibinfo {eid} {054107} (\bibinfo {year} {2012})}\BibitemShut
  {NoStop}%
\bibitem [{\citenamefont {Chakravorty}\ \emph {et~al.}(1993)\citenamefont
  {Chakravorty}, \citenamefont {Gwaltney}, \citenamefont {Davidson},
  \citenamefont {Parpia},\ and\ \citenamefont {p~Fischer}}]{CGDP93}%
  \BibitemOpen
  \bibfield  {author} {\bibinfo {author} {\bibfnamefont {S.~J.}\ \bibnamefont
  {Chakravorty}}, \bibinfo {author} {\bibfnamefont {S.~R.}\ \bibnamefont
  {Gwaltney}}, \bibinfo {author} {\bibfnamefont {E.~R.}\ \bibnamefont
  {Davidson}}, \bibinfo {author} {\bibfnamefont {F.~A.}\ \bibnamefont
  {Parpia}}, \ and\ \bibinfo {author} {\bibfnamefont {C.~F.}\ \bibnamefont
  {p~Fischer}},\ }\href {\doibase 10.1103/PhysRevA.47.3649} {\bibfield
  {journal} {\bibinfo  {journal} {Phys. Rev. A}\ }\textbf {\bibinfo {volume}
  {47}},\ \bibinfo {pages} {3649} (\bibinfo {year} {1993})}\BibitemShut
  {NoStop}%
\bibitem [{\citenamefont {Eshuis}, \citenamefont {Bates},\ and\ \citenamefont
  {Furche}(2012)}]{EBF12}%
  \BibitemOpen
  \bibfield  {author} {\bibinfo {author} {\bibfnamefont {H.}~\bibnamefont
  {Eshuis}}, \bibinfo {author} {\bibfnamefont {J.}~\bibnamefont {Bates}}, \
  and\ \bibinfo {author} {\bibfnamefont {F.}~\bibnamefont {Furche}},\ }\href
  {\doibase 10.1007/s00214-011-1084-8} {\bibfield  {journal} {\bibinfo
  {journal} {Theor. Chem. Acc.}\ }\textbf {\bibinfo {volume} {131}},\ \bibinfo
  {pages} {1} (\bibinfo {year} {2012})}\BibitemShut {NoStop}%
\bibitem [{\citenamefont {Leb\`egue}\ \emph {et~al.}(2010)\citenamefont
  {Leb\`egue}, \citenamefont {Harl}, \citenamefont {Gould}, \citenamefont
  {\'Angy\'an}, \citenamefont {Kresse},\ and\ \citenamefont
  {Dobson}}]{LHGAKD10}%
  \BibitemOpen
  \bibfield  {author} {\bibinfo {author} {\bibfnamefont {S.}~\bibnamefont
  {Leb\`egue}}, \bibinfo {author} {\bibfnamefont {J.}~\bibnamefont {Harl}},
  \bibinfo {author} {\bibfnamefont {T.}~\bibnamefont {Gould}}, \bibinfo
  {author} {\bibfnamefont {J.~G.}\ \bibnamefont {\'Angy\'an}}, \bibinfo
  {author} {\bibfnamefont {G.}~\bibnamefont {Kresse}}, \ and\ \bibinfo {author}
  {\bibfnamefont {J.~F.}\ \bibnamefont {Dobson}},\ }\href {\doibase
  10.1103/PhysRevLett.105.196401} {\bibfield  {journal} {\bibinfo  {journal}
  {Phys. Rev. Lett.}\ }\textbf {\bibinfo {volume} {105}},\ \bibinfo {pages}
  {196401} (\bibinfo {year} {2010})}\BibitemShut {NoStop}%
\bibitem [{\citenamefont {Bj\"orkman}\ \emph {et~al.}(2012)\citenamefont
  {Bj\"orkman}, \citenamefont {Gulans}, \citenamefont {Krasheninnikov},\ and\
  \citenamefont {Nieminen}}]{BGKN12}%
  \BibitemOpen
  \bibfield  {author} {\bibinfo {author} {\bibfnamefont {T.}~\bibnamefont
  {Bj\"orkman}}, \bibinfo {author} {\bibfnamefont {A.}~\bibnamefont {Gulans}},
  \bibinfo {author} {\bibfnamefont {A.~V.}\ \bibnamefont {Krasheninnikov}}, \
  and\ \bibinfo {author} {\bibfnamefont {R.~M.}\ \bibnamefont {Nieminen}},\
  }\href {http://stacks.iop.org/0953-8984/24/i=42/a=424218} {\bibfield
  {journal} {\bibinfo  {journal} {J. Phys.: Condens. Matter}\ }\textbf
  {\bibinfo {volume} {24}},\ \bibinfo {pages} {424218} (\bibinfo {year}
  {2012})}\BibitemShut {NoStop}%
\bibitem [{\citenamefont {Gould}\ and\ \citenamefont
  {Dobson}(2013{\natexlab{a}})}]{Gould2013-LEXX}%
  \BibitemOpen
  \bibfield  {author} {\bibinfo {author} {\bibfnamefont {T.}~\bibnamefont
  {Gould}}\ and\ \bibinfo {author} {\bibfnamefont {J.~F.}\ \bibnamefont
  {Dobson}},\ }\href {http://link.aip.org/link/?JCP/138/014103/1} {\bibfield
  {journal} {\bibinfo  {journal} {J. Chem. Phys.}\ }\textbf {\bibinfo {volume}
  {138}},\ \bibinfo {pages} {014103} (\bibinfo {year}
  {2013}{\natexlab{a}})}\BibitemShut {NoStop}%
\bibitem [{\citenamefont {Gould}\ and\ \citenamefont
  {Dobson}(2013{\natexlab{b}})}]{Gould2013-Aff}%
  \BibitemOpen
  \bibfield  {author} {\bibinfo {author} {\bibfnamefont {T.}~\bibnamefont
  {Gould}}\ and\ \bibinfo {author} {\bibfnamefont {J.~F.}\ \bibnamefont
  {Dobson}},\ }\href {http://link.aip.org/link/?JCP/138/014109/1} {\bibfield
  {journal} {\bibinfo  {journal} {J. Chem. Phys.}\ }\textbf {\bibinfo {volume}
  {138}},\ \bibinfo {pages} {014109} (\bibinfo {year}
  {2013}{\natexlab{b}})}\BibitemShut {NoStop}%
\bibitem [{\citenamefont {Gould}\ and\ \citenamefont {Toulouse}(2014)}]{GT14}%
  \BibitemOpen
  \bibfield  {author} {\bibinfo {author} {\bibfnamefont {T.}~\bibnamefont
  {Gould}}\ and\ \bibinfo {author} {\bibfnamefont {J.}~\bibnamefont
  {Toulouse}},\ }\href {\doibase 10.1103/PhysRevA.90.050502} {\bibfield
  {journal} {\bibinfo  {journal} {Phys. Rev. A}\ }\textbf {\bibinfo {volume}
  {90}},\ \bibinfo {pages} {050502} (\bibinfo {year} {2014})}\BibitemShut
  {NoStop}%
\bibitem [{\citenamefont {Chakravorty}\ and\ \citenamefont
  {Davidson}(1996)}]{CD96}%
  \BibitemOpen
  \bibfield  {author} {\bibinfo {author} {\bibfnamefont {S.}~\bibnamefont
  {Chakravorty}}\ and\ \bibinfo {author} {\bibfnamefont {E.}~\bibnamefont
  {Davidson}},\ }\href@noop {} {\bibfield  {journal} {\bibinfo  {journal} {J.
  Phys. Chem.}\ }\textbf {\bibinfo {volume} {100}},\ \bibinfo {pages} {6167}
  (\bibinfo {year} {1996})}\BibitemShut {NoStop}%
\bibitem [{\citenamefont {Gell-Mann}\ and\ \citenamefont
  {Brueckner}(1957)}]{GB57}%
  \BibitemOpen
  \bibfield  {author} {\bibinfo {author} {\bibfnamefont {M.}~\bibnamefont
  {Gell-Mann}}\ and\ \bibinfo {author} {\bibfnamefont {K.}~\bibnamefont
  {Brueckner}},\ }\href@noop {} {\bibfield  {journal} {\bibinfo  {journal}
  {Phys. Rev.}\ }\textbf {\bibinfo {volume} {106}},\ \bibinfo {pages} {364}
  (\bibinfo {year} {1957})}\BibitemShut {NoStop}%
\bibitem [{\citenamefont {{Onsager}}, \citenamefont {{Mittag}},\ and\
  \citenamefont {{Stephen}}(1966)}]{OMS66}%
  \BibitemOpen
  \bibfield  {author} {\bibinfo {author} {\bibfnamefont {L.}~\bibnamefont
  {{Onsager}}}, \bibinfo {author} {\bibfnamefont {L.}~\bibnamefont {{Mittag}}},
  \ and\ \bibinfo {author} {\bibfnamefont {M.~J.}\ \bibnamefont {{Stephen}}},\
  }\href@noop {} {\bibfield  {journal} {\bibinfo  {journal} {Ann. Phys.
  (Leipzig)}\ }\textbf {\bibinfo {volume} {18}} (\bibinfo {year}
  {1966})}\BibitemShut {NoStop}%
\bibitem [{foo({\natexlab{b}})}]{footnote2}%
  \BibitemOpen
  \bibinfo {note} {This may be derived from
  a least squares fit to a cubic polynomial of the nuclear charge Z as a
  function of $\nh$ for the noble gas atoms. This is inverted to an expression
  $\nh(Z)$ whose coefficients are integers nearly within fitting
  error.}\BibitemShut {Stop}%
\bibitem [{\citenamefont {Bates}\ and\ \citenamefont {Furche}(2013)}]{BF13}%
  \BibitemOpen
  \bibfield  {author} {\bibinfo {author} {\bibfnamefont {J.~E.}\ \bibnamefont
  {Bates}}\ and\ \bibinfo {author} {\bibfnamefont {F.}~\bibnamefont {Furche}},\
  }\href {\doibase 10.1063/1.4827254} {\bibfield  {journal} {\bibinfo
  {journal} {J. Chem. Phys.}\ }\textbf {\bibinfo {volume} {139}},\ \bibinfo
  {eid} {171103} (\bibinfo {year} {2013}),\ 10.1063/1.4827254}\BibitemShut
  {NoStop}%
\bibitem [{\citenamefont {Kurth}\ and\ \citenamefont
  {Perdew}(1999)}]{Kurth1999}%
  \BibitemOpen
  \bibfield  {author} {\bibinfo {author} {\bibfnamefont {S.}~\bibnamefont
  {Kurth}}\ and\ \bibinfo {author} {\bibfnamefont {J.~P.}\ \bibnamefont
  {Perdew}},\ }\href {\doibase 10.1103/PhysRevB.59.10461} {\bibfield  {journal}
  {\bibinfo  {journal} {Phys. Rev. B}\ }\textbf {\bibinfo {volume} {59}},\
  \bibinfo {pages} {10461} (\bibinfo {year} {1999})}\BibitemShut {NoStop}%
\bibitem [{\citenamefont {Yan}\ \emph {et~al.}(2000)\citenamefont {Yan},
  \citenamefont {Perdew}, \citenamefont {Kurth}, \citenamefont {Fiolhais},\
  and\ \citenamefont {Almeida}}]{Zidan2000}%
  \BibitemOpen
  \bibfield  {author} {\bibinfo {author} {\bibfnamefont {Z.}~\bibnamefont
  {Yan}}, \bibinfo {author} {\bibfnamefont {J.~P.}\ \bibnamefont {Perdew}},
  \bibinfo {author} {\bibfnamefont {S.}~\bibnamefont {Kurth}}, \bibinfo
  {author} {\bibfnamefont {C.}~\bibnamefont {Fiolhais}}, \ and\ \bibinfo
  {author} {\bibfnamefont {L.}~\bibnamefont {Almeida}},\ }\href {\doibase
  10.1103/PhysRevB.61.2595} {\bibfield  {journal} {\bibinfo  {journal} {Phys.
  Rev. B}\ }\textbf {\bibinfo {volume} {61}},\ \bibinfo {pages} {2595}
  (\bibinfo {year} {2000})}\BibitemShut {NoStop}%
\bibitem [{foo({\natexlab{c}})}]{footnote3}%
  \BibitemOpen
  \bibinfo {note} {Additional numerical
  tests suggest that our RPA energies may be systematically over-estimated,
  which would reduce the discrepancy betweeen theory and our calculations.
  However, as these possible systematic errors lie within the overall numerical
  error bars we do not attempt to correct for them.}\BibitemShut {Stop}%
\bibitem [{\citenamefont {Cangi}\ \emph {et~al.}(2010)\citenamefont {Cangi},
  \citenamefont {Lee}, \citenamefont {Elliott},\ and\ \citenamefont
  {Burke}}]{CLEB10}%
  \BibitemOpen
  \bibfield  {author} {\bibinfo {author} {\bibfnamefont {A.}~\bibnamefont
  {Cangi}}, \bibinfo {author} {\bibfnamefont {D.}~\bibnamefont {Lee}}, \bibinfo
  {author} {\bibfnamefont {P.}~\bibnamefont {Elliott}}, \ and\ \bibinfo
  {author} {\bibfnamefont {K.}~\bibnamefont {Burke}},\ }\href {\doibase
  10.1103/PhysRevB.81.235128} {\bibfield  {journal} {\bibinfo  {journal} {Phys.
  Rev. B}\ }\textbf {\bibinfo {volume} {81}},\ \bibinfo {pages} {235128}
  (\bibinfo {year} {2010})}\BibitemShut {NoStop}%
\bibitem [{\citenamefont {Cangi}\ \emph {et~al.}(2011)\citenamefont {Cangi},
  \citenamefont {Lee}, \citenamefont {Elliott}, \citenamefont {Burke},\ and\
  \citenamefont {Gross}}]{CLEB11}%
  \BibitemOpen
  \bibfield  {author} {\bibinfo {author} {\bibfnamefont {A.}~\bibnamefont
  {Cangi}}, \bibinfo {author} {\bibfnamefont {D.}~\bibnamefont {Lee}}, \bibinfo
  {author} {\bibfnamefont {P.}~\bibnamefont {Elliott}}, \bibinfo {author}
  {\bibfnamefont {K.}~\bibnamefont {Burke}}, \ and\ \bibinfo {author}
  {\bibfnamefont {E.~K.~U.}\ \bibnamefont {Gross}},\ }\href {\doibase
  10.1103/PhysRevLett.106.236404} {\bibfield  {journal} {\bibinfo  {journal}
  {Phys. Rev. Lett.}\ }\textbf {\bibinfo {volume} {106}},\ \bibinfo {pages}
  {236404} (\bibinfo {year} {2011})}\BibitemShut {NoStop}%
\bibitem [{\citenamefont {Constantin}\ \emph {et~al.}(2010)\citenamefont
  {Constantin}, \citenamefont {Snyder}, \citenamefont {Perdew},\ and\
  \citenamefont {Burke}}]{CSPB10}%
  \BibitemOpen
  \bibfield  {author} {\bibinfo {author} {\bibfnamefont {L.~A.}\ \bibnamefont
  {Constantin}}, \bibinfo {author} {\bibfnamefont {J.~C.}\ \bibnamefont
  {Snyder}}, \bibinfo {author} {\bibfnamefont {J.~P.}\ \bibnamefont {Perdew}},
  \ and\ \bibinfo {author} {\bibfnamefont {K.}~\bibnamefont {Burke}},\ }\href
  {\doibase 10.1063/1.3522767} {\bibfield  {journal} {\bibinfo  {journal} {J.
  Chem. Phys.}\ }\textbf {\bibinfo {volume} {133}},\ \bibinfo {pages} {241103}
  (\bibinfo {year} {2010})}\BibitemShut {NoStop}%
\bibitem [{\citenamefont {Csonka}, \citenamefont {Perdew},\ and\ \citenamefont
  {Ruzsinszky}(2010)}]{CPR10}%
  \BibitemOpen
  \bibfield  {author} {\bibinfo {author} {\bibfnamefont {G.~I.}\ \bibnamefont
  {Csonka}}, \bibinfo {author} {\bibfnamefont {J.~P.}\ \bibnamefont {Perdew}},
  \ and\ \bibinfo {author} {\bibfnamefont {A.}~\bibnamefont {Ruzsinszky}},\
  }\href {\doibase 10.1021/ct100488v} {\bibfield  {journal} {\bibinfo
  {journal} {J. Chem. Theory Comput.}\ }\textbf {\bibinfo {volume} {6}},\
  \bibinfo {pages} {3688} (\bibinfo {year} {2010})},\ \Eprint
  {http://arxiv.org/abs/http://dx.doi.org/10.1021/ct100488v}
  {http://dx.doi.org/10.1021/ct100488v} \BibitemShut {NoStop}%
\bibitem [{\citenamefont {R.~Bauernschmitt}(1996)}]{BA96}%
  \BibitemOpen
  \bibfield  {author} {\bibinfo {author} {\bibfnamefont {R.~A.}\ \bibnamefont
  {R.~Bauernschmitt}},\ }\href@noop {} {\bibfield  {journal} {\bibinfo
  {journal} {Chem. Phys. Lett.}\ }\textbf {\bibinfo {volume} {256}},\ \bibinfo
  {pages} {454} (\bibinfo {year} {1996})}\BibitemShut {NoStop}%
\bibitem [{\citenamefont {Buijse}, \citenamefont {Baerends},\ and\
  \citenamefont {Snijders}(1989)}]{BBS89}%
  \BibitemOpen
  \bibfield  {author} {\bibinfo {author} {\bibfnamefont {M.~A.}\ \bibnamefont
  {Buijse}}, \bibinfo {author} {\bibfnamefont {E.~J.}\ \bibnamefont
  {Baerends}}, \ and\ \bibinfo {author} {\bibfnamefont {J.~G.}\ \bibnamefont
  {Snijders}},\ }\href {\doibase 10.1103/PhysRevA.40.4190} {\bibfield
  {journal} {\bibinfo  {journal} {Phys. Rev. A}\ }\textbf {\bibinfo {volume}
  {40}},\ \bibinfo {pages} {4190} (\bibinfo {year} {1989})}\BibitemShut
  {NoStop}%
\bibitem [{\citenamefont {Burke}\ \emph {et~al.}(2014)\citenamefont {Burke},
  \citenamefont {Cancio}, \citenamefont {Gould},\ and\ \citenamefont
  {Pittalis}}]{BCGP14}%
  \BibitemOpen
  \bibfield  {author} {\bibinfo {author} {\bibfnamefont {K.}~\bibnamefont
  {Burke}}, \bibinfo {author} {\bibfnamefont {A.}~\bibnamefont {Cancio}},
  \bibinfo {author} {\bibfnamefont {T.}~\bibnamefont {Gould}}, \ and\ \bibinfo
  {author} {\bibfnamefont {S.}~\bibnamefont {Pittalis}},\ }\href
  {http://arxiv.org/abs/1409.4834v1} {\bibfield  {journal} {\bibinfo  {journal}
  {submitted and arXiv:1409.4834v1}\ } (\bibinfo {year} {2014})}\BibitemShut
  {NoStop}%
\bibitem [{\citenamefont {Sun}, \citenamefont {Ruzsinszky},\ and\ \citenamefont
  {Perdew}(2015)}]{SRP15}%
  \BibitemOpen
  \bibfield  {author} {\bibinfo {author} {\bibfnamefont {J.}~\bibnamefont
  {Sun}}, \bibinfo {author} {\bibfnamefont {A.}~\bibnamefont {Ruzsinszky}}, \
  and\ \bibinfo {author} {\bibfnamefont {J.~P.}\ \bibnamefont {Perdew}},\
  }\href {\doibase 10.1103/PhysRevLett.115.036402} {\bibfield  {journal}
  {\bibinfo  {journal} {Phys. Rev. Lett.}\ }\textbf {\bibinfo {volume} {115}},\
  \bibinfo {pages} {036402} (\bibinfo {year} {2015})}\BibitemShut {NoStop}%
\end{thebibliography}
%

\end{document}